\documentclass[12pt,epsfig]{article}
\usepackage{epsfig}

\begin{document}

\begin{center} 
{\Large {\bf Mass Spectrum in the Minimal Supersymmetric 3-3-1 model.}}
\end{center}

\begin{center}
M. C. Rodriguez  \\
{\it Universidade Federal do Rio Grande - FURG \\
Instituto de Matem\'atica, Estat\'\i stica e F\'\i sica - IMEF \\
Av. It\'alia, km 8, Campus Carreiros \\
96201-900, Rio Grande, RS \\
Brazil}
\end{center}

\date{\today}

\begin{abstract}
We consider the minimal supersymmetric extension of the 3-3-1 model. We 
study the mass spectra of this model in the fermionic and gauge bosons sectors 
without the antisextet. We also present some phenomenological consequences of 
this model at colliders such as Large Hadron Collider (LHC) and International 
Linear Collider(ILC).
\end{abstract}

PACS number(s): 12.60. Jv

Keywords: Supersymmetric models


\section{Introduction}

The models based on the $ \mbox{SU}(3)_C\otimes \mbox{SU}(3)_L
\otimes \mbox{U}(1)_X$ (called 3-3-1 models)
\cite{singer,ppf,331rh} provide possible solutions to some puzzles
of the Standard Model (SM) such as the generation number problem,
the electric charge quantization \cite{dongl2}. Since one
generation of quarks is treated differently from the others this
may  lead to a natural explanation for  the large mass of the top
quark~\cite{longvan}. There is also a good candidate for
SelfIinteracting Dark Matter (SIDM) since there are two Higgs
bosons, one scalar and one pseudoscalar, which have the properties
of candidates for dark matter like stability, neutrality and that
it must not overpopulate the universe~\cite{longlan}, etc.

There are two main versions of the 3-3-1 models depending on the
embedding of the charge operator in the $SU(3)_{L}$ generators,
\begin{equation}
\frac{Q}{e}= \frac{1}{2}(\lambda_3- \vartheta \lambda_8)+N \,\ I,
\label{co}
\end{equation}
where the $\vartheta$ parameter defines two different representation contents, 
$N$ denotes the $U(1)_{N}$ charge and $\lambda_3$, $\lambda_8$ are the diagonal 
generators of $SU(3)$.  In the minimal version, with $\vartheta=\sqrt3$,  the charge conjugation
of the right-handed charged lepton for each generation is combined
with the usual $SU(2)_L$ doublet left-handed leptons components to
form an $SU(3)$ triplet $(\nu, l, l^c)_L$.  No extra leptons are
needed in the mentioned  model, and we shall call such model as
minimal 3-3-1 model. There are also another possibility where the triplet 
$(\nu, l, L^c)_L$ where $L$ is an extra charged leptons wchich do not mix with the known 
leptons \cite{Pleitez:1992xh}. We want to remind that there is no right-handed (RH) neutrino in
both  model. There exists  another interesting possibility, where ($\vartheta=1/\sqrt3$) 
a left-handed anti-neutrino to each usual $SU(2)_L$ doublet is
added to form an $SU(3)$ triplet $(\nu, l , \nu^{c})_L$, and this model
is called the 3-3-1 model with RH neutrinos. The 3-3-1 models have
been studied extensively over the last decade.

The supersymmetric version of the 3-3-1 model minimal was done 
Refs.~\cite{331susy1,mcr} (MSUSY331) while the version with right-handed
neutrinos~\cite{331rh} has already been constructed in Ref.~\cite{331susy2,huong} (SUSY331RN), while 
the Supersymmetric economical 3-3-1 model with RH has been presented recently \cite{Dong:2007qc}(SUSYECO331).

Recently we have already constructed all the spectrum from the scalar sector from the MSUSY331 model 
\cite{Rodriguez:2005jt}. All the results obtained on that article are in agreement with the 
experimental limits. On this article we want to present the results about the masses in the fermion's 
sector and in the gauge boson's sector.

This paper is organized as follows. In Sec. \ref{psusy1} we review
the minimal supersymmetric 331 model while in Sec. \ref{sec:rparity} we show how we can define one $R$-parity in our model 
such that the neutrino's get their masses and keeping the proton's safe. While in Sections (\ref{sec:pheno}) 
and (\ref{sec:feno2}) we present some phenomenological consequences of this model to the colliders physics. While on 
Sec. (\ref{sec:massspectrum}) we present the mass values of all the fermions and gauge bosons of this model.  Finally, 
the last section is devoted to our conclusions. In Appendix \ref{sec:lagrangian} we present the Lagrangian of 
this model in terms of the fields.

\section{Minimal Supersymmetric 3-3-1 model (MSUSY331).}
\label{psusy1}

On this Section, we present our model. We start to introduce the minimal set of particle necessary to get the 
supersymmetric version of model given at Ref.~\cite{Pleitez:1992xh}. After the introduction of the particle content 
of our model we put them in the superfields, see Sec.(\ref{subsec:superfields}), and then we construct the full 
Lagrangian of our model in Sec.(\ref{lagrangian}). Then we show the pattern of the symmetry breaking of the model at Sec.(\ref{breaksusy331}).

\subsection{Particle Content}
\label{pcontent}

In the nonsupersymmetric 3-3-1 model~\cite{ppf} the fermionic representation
content is as follows: left-handed leptons 
$L_{aL}=(\nu_{a},l_{a},l^{c}_{a})_{L}\sim({\bf1},{\bf3},0)$, 
$a=e,\mu,\tau$; left-handed quarks 
$Q_{\alpha L}=(d_{\alpha},u_{\alpha},j_{\alpha})_{L}\sim({\bf3},{\bf3}^*,-1/3)$
, $\alpha=1,2$, $Q_{3L}=(u_{3},d_{3},J)_{L}\sim({\bf3},{\bf3},2/3)$; and in 
the right-handed components we have the usual quarks $u^{c}_{iL}\sim({\bf3}^{*},{\bf1},-2/3),
d^{c}_{iL}\sim({\bf3}^{3},{\bf1},1/3),\,i=1,2,3$, and the exotic quarks 
$j^{c}_{\alpha L}\sim({\bf3}^{*},{\bf1},4/3),J^{c}_{L}\sim({\bf3}^{*},{\bf1},-5/3)$, they have charge 
$-(4/3)e$ and $(5/3)e$ respectivelly. The minimal scalar representation content is formed by three scalar triplets: 
$\eta\sim({\bf1},{\bf3},0)=(\eta^{0},\eta^{-}_{1},\eta^{+}_{2})^T$;
$\rho\sim({\bf1},{\bf3},+1)=(\rho^{+}, \rho^{0}, \rho^{++})^T$ and 
$\chi\sim({\bf1},{\bf3},-1)=(\chi^{-},\chi^{--},\chi^{0})^T$.

Now, we introduce the minimal set of particles in order to implement the
supersymmetry~\cite{haber}. We have to introduce the sleptons the superpartners of the leptons and the 
squarks related to the quarks, both are scalars. Therefore in the supersymmetric version of this model 
\cite{331susy1,mcr}, the fermionic content is the same as in the nonsupersymmetric 331 model and we have to add their 
supersymmetric partners $\tilde{L}_{aL}$, $\tilde{Q}_{\alpha L}$, $\tilde{Q}_{3L}$, $\tilde{u}^{c}_{iL}$, 
$\tilde{d}^{c}_{iL}$, $\tilde{j}^{c}_{\alpha L}$ and $\tilde{J}^{c}_{L}$. We have also to introduce the higgsinos 
the supersymmetric partner of the scalars of the model and the minimal higgsinos are given by $\tilde{\eta}$, 
$\tilde{\rho}$ and $\tilde{\chi}$. However, we have to introduce, the 
followings extras scalars $\eta^{\prime}$, $\rho^{\prime}$, $\chi^{\prime}$ and their higgsinos 
$\tilde{\eta}^{\prime}$, $\tilde{\rho}^{\prime}$ and $\tilde{\chi}^{\prime}$, in order to to cancel chiral 
anomalies generated by $\tilde{\eta}$, $\tilde{\rho}$ and $\tilde{\chi}$.

Concerning the gauge bosons and their superpartners the gauginos. We denote
the gluons by $g^b$, the respective superparticles, the gluinos,
are denoted by $\lambda^b_{C}$, with $b=1, \ldots,8$; and in the
electroweak sector we have $V^b$, the gauge boson of $SU(3)_{L}$,
and their gauginos  $\lambda^b_{A}$; finally we have the gauge
boson of $U(1)_{N}$, denoted by $V^{\prime}$, and its
supersymmetric partner $\lambda_{B}$.

This is the minimal number of fields in the minimal supersymmetric
extension of the 3-3-1 model of Refs.~\cite{331susy1,mcr}. Summaryzing, 
we have in the 3-3-1 supersymmetric model the following superfields:
$\hat{L}_{e,\mu,\tau}$, $\hat{Q}_{1,2,3}$, $\hat{\eta}$, $\hat{\rho}$, 
$\hat{\chi}$; $\hat{\eta}^\prime$, $\hat{\rho}^\prime$, 
$\hat{\chi}^\prime$; $\hat{u}^c_{1,2,3}$, $\hat{d}^c_{1,2,3}$, $\hat{J}$ and $\hat{j}_{1,2}$, i.e., 21 chiral
superfields, and 17 vector superfields: $\hat{V}^a$, $\hat{V}^\alpha$ and
$\hat{V}^\prime$. In the Minimal Supersymmetric Standard Model (MSSM) 
~\cite{haber,Fayet:1974pd,Fayet:1976et,Fayet:1977yc,Fayet:1977vd,fayet78rp,Dawson,dress,tata} there are
14 chiral superfields and 12 vector superfields.

\subsection{Superfields}
\label{subsec:superfields}

The superfields formalism is useful in writing the Lagrangian which is 
manifestly invariant under the supersymmetric transformations~\cite{wb} with 
fermions and scalars put in chiral superfields while the gauge bosons in 
vector superfields. As usual the superfield of a field $\phi$ will be denoted
by $\hat{\phi}$~\cite{haber}.
The chiral superfield of a multiplet $\phi$ is denoted by~\cite{wb} 
\begin{eqnarray}
\hat{\phi}\equiv\hat{\phi}(x,\theta,\bar{\theta})&=& \tilde{\phi}(x) 
+ i \; \theta \sigma^{m} \bar{ \theta} \; \partial_{m} \tilde{\phi}(x) 
+\frac{1}{4} \; \theta \theta \; \bar{ \theta}\bar{ \theta} \; \partial^{m}\partial_{m} 
\tilde{\phi}(x) \nonumber \\ 
& & \mbox{} +  \sqrt{2} \; \theta \phi(x) 
+ \frac{i}{ \sqrt{2}} \; \theta \theta \; \bar{ \theta} \bar{ \sigma}^{m}
\partial_{m}\phi(x)                   
+  \theta \theta \; F_{\phi}(x), 
\label{phi}
\end{eqnarray}
while the vector superfield is given by
\begin{eqnarray}
\hat{V}(x,\theta,\bar\theta)&=&-\theta\sigma^m\bar\theta V_m(x)
+i\theta\theta\bar\theta
\overline{\tilde{V}}(x)-i\bar\theta\bar\theta\theta
\tilde{V}(x) 
+ \frac{1}{2}\theta\theta\bar\theta\bar\theta D(x).
\label{vector}
\end{eqnarray}
The fields $F$ and $D$ are auxiliary fields which are needed to
close the supersymmetric algebra and eventually will be eliminated using 
their motion equations.

In the nonsupersymmetric 3-3-1 model to give arbitrary mass to the leptons we 
have to introduce one scalar 
antisextet $S\sim({\bf1},{\bf6}^{*},0)$.We can avoid the introduction of the 
antisextet by adding a charged lepton transforming as a singlet. 
Notwithstanding, here we will omit both the antisextet, we are going to show in 
Sec.\ref{sec:massspectrum} all the fermions and gauge bosons of the model get 
their masses only with three triplets and three antitriplets in agreement with 
\cite{lepmass}. 

\subsection{The Lagrangian}
\label{lagrangian}

On this subsection we will write only the lagrangian in the terms of superfields of the model.
The Lagrangian of the model has the following form 
\begin{eqnarray} 
   {\cal L}_{331S} &=& {\cal L}_{SUSY} + {\cal L}_{\mbox{soft}}, 
\label{lagra}
\end{eqnarray}
where ${\cal L}_{SUSY}$ is the supersymmetric part and  
${\cal L}_{\mbox{soft}}$ the soft terms breaking explicitly the supersymmetry.

The supersymmetric term can be divided as follows
\begin{equation} 
   {\cal L}_{SUSY} =   {\cal L}_{Lepton}
                  + {\cal L}_{Quarks} 
                  + {\cal L}_{Gauge} 
                  + {\cal L}_{Scalar}, 
\label{l2}
\end{equation}
The fermion's lagrangian is given by
\begin{eqnarray}
{\cal L}_{\mbox{Lepton}} 
&=& \int d^{4}\theta\;\left[\,\hat{ \bar{L}}_{a}e^{2g\hat{V}} \hat{L}_{a} \,\right], \nonumber \\
{\cal L}_{\mbox{Quarks}} 
&=& \int d^{4}\theta\;\left \{ 
\,\hat{\bar{Q}}_{\alpha}e^{ \left[ 2g_{s}\hat{V}_{C}+2g\hat{\bar{V}}+g^{\prime}
\left( - \frac{1}{3} \right) \hat{V}^{\prime} \right]} \hat{Q}_{\alpha}
+\,\hat{ \bar{Q}}_{3}e^{ \left[ 2g_{s}\hat{V}_{C}+2g\hat{V}+g^{\prime}
\left( \frac{2}{3}\right) \hat{V}^{\prime} \right]} \hat{Q}_{3} \right. \nonumber \\
&+&\left. 
\hat{ \bar{u}}^{c}_{i}e^{ \left[ 2g_{s} \hat{ \bar{V}}_{C}+g^{\prime}
\left( - \frac{2}{3}\right) \hat{V}^{\prime} \right]} \hat{u}^{c}_{i}+ 
\hat{ \bar{d}}^{c}_{i}e^{ \left[ 2g_{s} \hat{ \bar{V}}_{C}+g^{\prime} 
\left( \frac{1}{3}\right)\hat{V}^{\prime} \right]} \hat{d}^{c}_{i} 
+ \hat{ \bar{J}}^{c}e^{ \left[ 2g_{s} \hat{ \bar{V}}_{C}+g^{\prime}
\left( - \frac{5}{3}\right)\hat{V}^{\prime} \right]} \hat{J}^{c} \right. \nonumber \\ 
&+&\left. 
\hat{ \bar{j}}^{c}_{ \alpha}e^{ \left[ 2g_{s} \hat{ \bar{V}}_{C}+g^{\prime}
\left( \frac{4}{3}\right)\hat{V}^{\prime} \right]} \hat{j}^{c}_{ \alpha} \right\} .
\end{eqnarray}
where we have defined $\hat{V}_C=T^b\hat{V}^b_C$, 
$\hat{V}=T^b\hat{V}^b$; $\hat{\bar V}_C=\bar{T}^b\hat{V^b}_C$,
$\hat{\bar V}=\bar{T}^b\hat{V^b}$;
$T^b=\lambda^b/2$, $\bar{T}^b=-\lambda^{*b}/2$ are the generators of triplet
and antitriplets representations, respectively, and $\lambda^b$ are the
Gell-Mann matrices.  

In the gauge sector we have
\begin{eqnarray}
{\cal L}_{gauge} 
&=&  \frac{1}{4} \int  d^{2}\theta\;Tr[ {\cal W}_{C} {\cal W}_{C}]+ \frac{1}{4} 
\int  d^{2}\theta\;Tr[ {\cal W}_{L} {\cal W}_{L}]+
\frac{1}{4} \int  d^{2}\theta {\cal W}^{ \prime} {\cal W}^{ \prime} \nonumber \\ &+&  
\frac{1}{4} \int  d^{2}\bar{\theta}\;Tr[\bar{{\cal W}}_{C}\bar{{\cal W}}_{C}]+ \frac{1}{4} 
\int  d^{2}\bar{\theta}\;Tr[\bar{{\cal W}}_{L}\bar{{\cal W}}_{L}]+
\frac{1}{4} \int  d^{2}\bar{\theta} \bar{{\cal W}}^{ \prime}\bar{{\cal W}}^{ \prime}\,\ , 
\nonumber \\
\label{gaugm1}
\end{eqnarray}
where ${\cal W}_{C}$, ${\cal W}_{L}$ e ${\cal W}^{ \prime}$ are fields that can be
written as follows~\cite{wb} 
\begin{eqnarray}
{\cal W}_{\zeta C}&=&- \frac{1}{8g} \bar{D} \bar{D} e^{-2g \hat{V}_{C}} 
D_{\zeta} e^{2g_s \hat{V}_{C}},\nonumber \\ 
{\cal W}_{\zeta L}&=&- \frac{1}{8g} \bar{D} \bar{D} e^{-2g \hat{V}} 
D_{\zeta} e^{2g \hat{V}}, \nonumber \\
{\cal W}^{\prime}_{\zeta}&=&- \frac{1}{4} \bar{D} \bar{D} D_{\zeta} 
\hat{V}^{\prime}, \,\ \zeta=1,2.
\label{cforca}
\end{eqnarray}

Finally, in the scalar sector we have
\begin{eqnarray}
{\cal L}_{scalar} 
&=& \int d^{4}\theta\;\left[\,\hat{ \bar{ \eta}}e^{2g\hat{V}} \hat{ \eta} + 
\hat{ \bar{ \rho}}e^{ \left( 2g\hat{V}+g^{\prime}\hat{V}^{\prime} \right)} \hat{ \rho} +
\hat{ \bar{ \chi}}e^{ \left( 2g\hat{V}-g^{\prime}\hat{V}^{\prime} \right)} \hat{ \chi} \right. \nonumber \\
&+& \left.\,\hat{ \bar{ \eta}}^{\prime}e^{2g\hat{ \bar{V}}} \hat{ \eta}^{\prime} + 
\hat{ \bar{ \rho}}^{\prime}e^{ \left( 2g\hat{ \bar{V}}-g^{\prime}\hat{V}^{\prime} \right)} \hat{ \rho}^{\prime} +
\hat{ \bar{ \chi}}^{\prime}e^{ \left( 2g\hat{ \bar{V}}+g^{\prime}\hat{V}^{\prime} \right)} \hat{ \chi}^{\prime} \right]
+ \int d^{2}\theta W+ \int d^{2}\bar{ \theta}\overline{W} \,\ , \nonumber \\
\label{escm1}
\end{eqnarray}
here $g$ and $g^{\prime}$ are the gauge coupling constants of $SU(3)$ and 
$U(1)$ respectivelly and $W$ is the superpotential of the model.

The superpotential of our model is given by
\begin{equation}
W=\frac{W_{2}+ \bar{W}_{2}}{2}+ \frac{W_{3}+ \bar{W}_{3}}{3}, 
\label{sp1}
\end{equation}
with $W_{2}$ having only two chiral superfields while $W_{3}$ has three chiral superfields. The terms allowed by  
our symmetry are
\begin{eqnarray}
W_{2}&=&\mu_{0a}\hat{L}_{aL} \hat{ \eta}^{\prime}+ 
\mu_{ \eta} \hat{ \eta} \hat{ \eta}^{\prime}+
 \mu_{ \rho} \hat{ \rho} \hat{ \rho}^{\prime}+ 
\mu_{ \chi} \hat{ \chi} \hat{ \chi}^{\prime}, \nonumber \\
W_{3}&=& \lambda_{1abc} \epsilon \hat{L}_{aL} \hat{L}_{bL} \hat{L}_{cL}+
\lambda_{2ab} \epsilon \hat{L}_{aL} \hat{L}_{bL} \hat{ \eta}+ 
\lambda_{3a} \epsilon 
\hat{L}_{aL} \hat{\chi} \hat{\rho}+
f_{1} \epsilon \hat{ \rho} \hat{ \chi} \hat{ \eta}+
f^{\prime}_{1}\epsilon \hat{ \rho}^{\prime}\hat{ \chi}^{\prime}\hat{ \eta}^{\prime} \nonumber \\
&+&
\kappa_{1\alpha i} \hat{Q}_{\alpha L} \hat{\rho} \hat{u}^{c}_{iL} +  
\kappa_{2\alpha i} \hat{Q}_{\alpha L} \hat{\eta} \hat{d}^{c}_{iL}+
\kappa_{3\alpha \beta} \hat{Q}_{\alpha L} \hat{\chi} \hat{j}^{c}_{\beta L} \nonumber \\
&+& 
\kappa_{4\alpha ai} \hat{Q}_{\alpha L} \hat{L}_{aL} \hat{d}^{c}_{iL}+ 
\kappa_{5i} \hat{Q}_{3L} \hat{\eta}^{\prime} \hat{u}^{c}_{iL}+
\kappa_{6i} \hat{Q}_{3L} \hat{\rho}^{\prime} \hat{d}^{c}_{iL}+
\kappa_{7} \hat{Q}_{3L} \hat{\chi}^{\prime} \hat{J}^{c}_{L} \nonumber \\
&+&
\xi_{1ijk} \hat{d}^{c}_{iL} \hat{d}^{c}_{jL} \hat{u}^{c}_{kL}+
\xi_{2ij \beta} \hat{u}^{c}_{iL} \hat{u}^{c}_{jL} \hat{j}^{c}_{\beta L}+
\xi_{3i \beta} \hat{d}^{c}_{iL} \hat{J}^{c}_{L} \hat{j}^{c}_{\beta L}. 
\label{sp3m1}
\end{eqnarray}
The coefficients $\mu_{0}, \mu_{\eta}, \mu_{\rho}$ and $\mu_{\chi}$ have mass 
dimension, while all the coefficients in $W_{3}$ are dimensionless \cite{dress,tata}. To see the lagrangian of 
this model in terms of the fields see Appendix \ref{sec:lagrangian}.

The most general soft supersymmetry breaking terms,
which do not induce quadratic divergence, where described by
Girardello and Grisaru \cite{10}. They found that the allowed
terms can be categorized as follows: 
\begin{itemize}
\item scalar mass term
\begin{equation}
{\cal L}_{SMT}=-m^{2} A^{\dagger}A,
\end{equation}
\item gaugino mass  term
\begin{equation}
{\cal L}_{GMT}=- \frac{1}{2} (M_{ \lambda} \lambda^{a} \lambda^{a}+H.c)
\end{equation}
\item scalar interaction terms
\end{itemize}
\begin{equation}
{\cal L}_{int}= m_{ij}A_{i}A_{j}+ f_{ijk}\epsilon^{ijk}A_{i}A_{j}A_{k}+H.c.
\end{equation}
The terms on this case are similar with the terms allowed in the
superpotential of the model we are considering, see Eq.(\ref{sp3m1}).

They, also, must be consistent with the 3-3-1 gauge symmetry. These soft terms are given by
\begin{equation}
{\cal L}_{\mbox{soft}}={\cal L}_{GMT}+
{\cal L}^{\mbox{soft}}_{\mbox{scalar}}+{\cal L}_{SMT},
\end{equation}
where
\begin{eqnarray}
{\cal L}_{GMT}=- \frac{1}{2} \left[m_{ \lambda_{C}} \sum_{a=1}^{8} 
\left( \lambda^{a}_{C} \lambda^{a}_{C} \right) +m_{ \lambda} \sum_{a=1}^{8} 
\left( \lambda^{a}_{A} \lambda^{a}_{A} \right) 
+m^{ \prime} \lambda_{B} \lambda_{B}+H.c. \right],
\label{gmt}
\end{eqnarray}
due this term the gauginos get theis masses at scale where SUSY is broken while their 
superpartners the gauge bosons are massless, for more detail about symmetry breaking inthis model 
see Sec.(\ref{breaksusy331}). The second term give masses to the higgsinos is written as
\begin{eqnarray}
{\cal L}_{SMT}&=&
-m^2_{ \eta}\eta^{ \dagger}\eta-m^2_{ \rho}\rho^{ \dagger}\rho-
m^2_{ \chi}\chi^{ \dagger}\chi
-m^2_{\eta^{\prime}}\eta^{\prime \dagger}\eta^{\prime}-
m^2_{\rho^{\prime}}\rho^{\prime \dagger}\rho^{\prime}-
m^2_{\chi^{\prime}}\chi^{\prime \dagger}\chi^{\prime} \nonumber \\
&-&m_{L}^{2} \tilde{L}^{\dagger}_{aL} \tilde{L}_{aL}-
m_{Q_{\alpha}}^{2} \tilde{Q}^{\dagger}_{\alpha L} \tilde{Q}_{\alpha L}-
m_{Q_3}^{2} \tilde{Q}^{\dagger}_{3L} \tilde{Q}_{3L}-
m_{u_{i}}^2 \tilde{u}^{c \dagger}_{iL} \tilde{u}^{c}_{iL}- 
m_{d_{i}}^2 \tilde{d}^{c \dagger}_{iL} \tilde{d}^{c}_{iL}-
m_{J}^{2} \tilde{J}^{c \dagger}_{L} \tilde{J}^{c}_{L} \nonumber \\
&-&m_{j_{ \beta}}^{2} \tilde{j}^{c \dagger}_{ \beta L} \tilde{j}^{c}_{ \beta L}
+[k_1\epsilon \rho \chi \eta+
k^{\prime}_1\epsilon \rho^{\prime} \chi^{\prime} \eta^{\prime}+
H.c.], 
\end{eqnarray}
while the last term is given by
\begin{eqnarray}
{\cal L}_{int}&=& \left[-M_{a}^2 \tilde{L}_{aL} \eta^{\dagger}+
\varepsilon_{0abc}  \epsilon \tilde{L}_{aL} \tilde{L}_{bL} \tilde{L}_{cL}+
\varepsilon_{1ab} \epsilon \tilde{L}_{aL} \tilde{L}_{bL} \eta +
\varepsilon_{2a} \epsilon \tilde{L}_{aL} \chi \rho  \right. \nonumber \\
&+& \left.
\tilde{Q}_{\alpha L} \left( \omega_{1\alpha i} \eta 
\tilde{d}^{c}_{iL} 
+ \omega_{2\alpha i} \rho \tilde{u}^{c}_{iL}+
\omega_{3 \alpha aj}  \tilde{L}_{aL} \tilde{d}^{c}_{jL}
+\omega_{4\alpha \beta}  \chi \tilde{j}^{c}_{\beta L} \right) \right. \nonumber \\
&+& \left.
\tilde{Q}_{3L}( \zeta_{1i}  \eta^{\prime} \tilde{u}^{c}_{iL}+
\zeta_{2i}  \rho^{\prime} \tilde{d}^{c}_{iL}+ \zeta_{3J} 
\chi^{\prime} \tilde{J}^{c}_{L}) + 
\varsigma_{1ijk} \tilde{d}^{c}_{iL} \tilde{d}^{c}_{jL} \tilde{u}^{c}_{kL}+ 
\varsigma_{2i\beta} \tilde{d}^{c}_{iL} \tilde{J}^{c}_{L} \tilde{j}^{c}_{\beta L} \right. \nonumber \\
&+& \left.
\varsigma_{3ij\beta} \tilde{u}^{c}_{iL} \tilde{u}^{c}_{jL} \tilde{j}^{c}_{\beta L}+H.c.
\right]. 
\end{eqnarray}

\subsection{Breake structure from MSUSY331 to $SU(3)_{C} \otimes U(1)_{Q}$}
\label{breaksusy331}

The pattern of the symmetry breaking of the model is given by the following 
scheme(using the notation given at \cite{ppf})
\begin{eqnarray}
&\mbox{MSUSY331}&
\stackrel{{\cal L}_{soft}}{\longmapsto}
\mbox{SU(3)}_C\ \otimes \ \mbox{SU(3)}_{L}\otimes \mbox{U(1)}_{N}
\stackrel{\langle\chi\rangle \langle \chi^{\prime}\rangle}{\longmapsto}
\mbox{SU(3)}_{C} \ \otimes \ \mbox{SU(2)}_{L}\otimes
\mbox{U(1)}_{Y} \nonumber \\
&\stackrel{\langle\rho,\eta, \rho^{\prime}\eta^{\prime}\rangle}{\longmapsto}&
\mbox{SU(3)}_{C} \ \otimes \ \mbox{U(1)}_{Q}
\label{breaksusy331tou1}
\end{eqnarray}

When one breaks the 3-3-1 symmetry to the $SU(3)_{C} \otimes U(1)_{Q}$, the 
scalars get the following vacuum expectation values (VEVs):
\begin{eqnarray} 
< \eta > &=& 
      \left( \begin{array}{c} v \\ 
                  0 \\
                  0          \end{array} \right),\quad 
< \rho > = 
      \left( \begin{array}{c} 0 \\ 
                  u \\
                  0          \end{array} \right),\quad 
< \chi > = 
      \left( \begin{array}{c} 0 \\ 
                  0 \\
                  w          \end{array} \right), \nonumber \\
< \eta^{\prime} > &=& 
      \left( \begin{array}{c} v^{\prime} \\ 
                  0 \\
                  0          \end{array} \right),\quad 
< \rho^{\prime} > = 
      \left( \begin{array}{c} 0 \\ 
                  u^{\prime} \\
                  0          \end{array} \right),\quad 
< \chi^{\prime} > = 
      \left( \begin{array}{c} 0 \\ 
                  0 \\
                  w^{\prime}          \end{array} \right), 
\label{vev1} 
\end{eqnarray}
where $v=v_{\eta}/ \sqrt{2}$, $u=v_{\rho}/ \sqrt{2}$, 
$w=v_{\chi}/ \sqrt{2}$, $v^{\prime}=v_{\eta^{\prime}}/ \sqrt{2}$, 
$u^{\prime}=v_{\rho^{\prime}}/ \sqrt{2}$ and 
$w^{\prime}=v_{\chi^{\prime}}/ \sqrt{2}$. From this pattern of the symmetry breaking comes the following 
constraint \cite{mcr}
\begin{equation} 
V^{2}_{\eta}+V^{2}_{\rho}=(246\;{\rm GeV})^2
\label{wmasslimite}
\end{equation} 
coming from $M_W$, where, we have defined $V^{2}_{\eta}=v^{2}_{\eta}+v^{\prime 2}_{\eta}$ 
and $V^{2}_{\rho}= v^{2}_{\rho}+v^{\prime 2}_{\rho}$. Therefore the VEV's of our model satisfy the conditions:
\begin{equation}
w,w^{\prime} \gg v,v^{\prime}, u,u{^\prime}.\label{cond}
\end{equation}

\section{Phenomenological Consequences in the lepton's and quark's sectors.}
\label{sec:pheno}

In the usual 3-3-1 model \cite{ppf} the gauge bosons are defined as
\begin{eqnarray}
W^{ \pm}_{m}(x)&=&-\frac{1}{\sqrt{2}}(V^{1}_{m}(x) \mp i V^{2}_{m}(x)),
\,\
V^{ \pm}_{m}(x)=-\frac{1}{\sqrt{2}}(V^{4}_{m}(x) \pm i V^{5}_{m}(x)), 
\nonumber \\
U^{\pm \pm}_{m}(x) &=&- \frac{1}{\sqrt{2}}(V^{6}_{m}(x) \pm i V^{7}_{m}(x)), 
\,\
A_{m}(x) = \frac{1}{\sqrt{1+4t^{2}}}
\left[ (V^{3}_{m}(x)- \sqrt{3}V^{8}_{m}(x))t+V_{m} \right], 
\nonumber \\
Z^{0}_{m}(x)  &=&- \frac{1}{\sqrt{1+4t^{2}}}
\left[ \sqrt{1+3t^{2}}V^{3}_{m}(x)+ 
\frac{ \sqrt{3}t^{2}}{\sqrt{1+3t^{2}}}V^{8}_{m}(x)-
\frac{t}{\sqrt{1+3t^{2}}}V_{m}(x) \right], \nonumber \\
Z^{\prime 0}_{m}(x) &=& \frac{1}{\sqrt{1+3t^{2}}}
(V^{8}_{m}(x)+ \sqrt{3}tV_{m}(x)),
\label{defbosons}
\end{eqnarray}
where $t \equiv \tan \theta = \frac{g^{ \prime}}{g}$ and $g^{\prime}$ and $g$ 
are the gauge coupling constants of $U(1)$ and $SU(3)$, 
respectively. 

The bosons $U^{--}$ and $V^{-}$ are called bileptons because they couple to 
two leptons; thus they have two units of lepton number, it means $L=2$. Here 
$L$ is the total lepton number, give by $L=L_{e}+L_{\mu}+L_{\tau}$. This model 
does not conserve separate family lepton number, $L_{e}$, $L_{\mu}$ and 
$L_{\tau}$ but only the total lepton number $L$ is conserved.

We can define the charged gauginos, in analogy with the gauge bosons in the MSSM, in the following way \cite{mcr}
\begin{eqnarray}
\lambda_{W}^{\pm}(x)&=& -\frac{1}{\sqrt{2}}(\lambda^{1}_{A}(x) \mp i 
\lambda^{2}_{A}(x)), \,\
\lambda_{V}^{\pm}(x)=-\frac{1}{\sqrt{2}}(\lambda^{4}_{A}(x) \pm i 
\lambda^{5}_{A}(x)),\nonumber \\
\lambda_{U}^{\pm \pm}(x) &=&-\frac{1}{\sqrt{2}}(\lambda^{6}_{A}(x) \pm i 
\lambda^{7}_{A}(x)).
\label{defgauginos}
\end{eqnarray}

The charged current interactions for the fermions, came from ${\cal L}_{llV}$ (first equation at Eq.(\ref{lepbos})) and 
from ${\cal L}_{qqV}$ (first equation at Eq.(\ref{quarkbos})) we can rewrite them in the following way
\begin{eqnarray}
{\cal L}_l^{CC}&=&-\frac{g}{\sqrt2}\sum_l\left(\bar\nu_{lL}\gamma^m
l_LW^+_m+ \bar l^c_L\gamma^m\nu_{lL} V^+_m+\bar l^c_L\gamma^m l_L
U^{++}_m+h.c.\right), \nonumber \\
{\cal L}_q^{CC}& = & -\frac{g}{2\sqrt{2}}\left[\overline{U}\gamma^m(1 - 
\gamma_5)V_{\rm CKM}DW^+_m  + \overline{U}\gamma^m(1 - \gamma_5)\zeta{\cal 
JV}_m + \overline{D}\gamma^m(1 - \gamma_5)\xi{\cal JU}_m\right] \nonumber \\
&+& {\mbox {\rm H. c.}}, 
\nonumber \\
\label{lq}
\end{eqnarray}
where we have defined the mass eigenstates in the following way
\begin{eqnarray}
U = \left(\begin{array}{c} u \\ c \\ t
\end{array}\right), \quad
D & = & \left(\begin{array}{c}
         d \\ s \\ b
\end{array}\right), \quad
{\cal V}_m = \left(\begin{array}{c}
         V^+_m \\ U^{--}_m \\ U^{--}_m\end{array}\right), \nonumber \\
{\cal U}_m & = & \left(\begin{array}{c}
         U^{--}_m \\ V^+_m \\ V^+_m
\end{array}\right), 
\end{eqnarray}\label{maest}
and ${\cal J} = {\rm diag}\left(\begin{array}{ccc}J_1 & J_2 & 
J_3\end{array}\right)$. 
The $V_{\rm CKM}$ is the usual Cabibbo-Kobayashi-Maskawa mixing matrix and $\xi$ 
and $\zeta$ are mixing matrices containing new unknown mixing parameters due to 
the presence of the exotic quarks. 

We can calculate the Higgs couplings to the usual leptons on this model is given by ${\cal L}_{llH}$, 
see Eq.(\ref{HMT}) at Appendix \ref{sec:lagrangian}, we get the following lagrangian \cite{ppf}
\begin{eqnarray}
{\cal L}_{llH}&=& \frac{\lambda_{2ab}}{3}(-\bar l_{aR}l_{bL}\eta^0+\bar l_{aR}\nu_{bL}\eta^-_1 +
\bar\nu^{c}_{aR}l_{aL}\eta^+_2+ H.c.),
\label{higgslept} 
\end{eqnarray}
the coupling ${\cal L}_{qqH}$ is the same as get in \cite{ppf} on their Eq.(13). On this model the same flavor leptons 
they don't couple with the neutral Higgs, therefore our lighest Higgs doesn't couples with two electrons 
\footnote{I would like to thanks E. Gregores that call my attention to this dangeours Higgs decay channel.} and of course 
it can decay in the following way $H^{0}_{1}\to e^{\pm}\mu^{\mp}$, and their coupling is $\lambda_{2e \mu}=10^{-3}$, 
see \cite{lepmass}, due this fact our light Higgs with $m_{H^{0}_{1}}=110,5$GeV was not detected \cite{Rodriguez:2005jt}
 by the experiment Large Electron Positron (LEP).

We have already showed that in the M\o ller scattering and in 
muon-muon scattering we can show that left-right asymmetries $A_{RL}(ll)$  are very sensitive to a 
doubly charged vector bilepton resonance but they are insensitive to scalar ones 
\cite{Montero:1998ve,Montero:1999en,Montero:2000ch}.

Similarly, we have the neutral currents coupled to both $Z^0$ and
$Z'^0$ massive vector bosons, according to the Lagrangian
\begin{equation}
{\cal L}_\nu^{NC}=-\frac{g}{2}\frac{M_Z}{M_W}
\bar\nu_{lL}\gamma^m\nu_{lL}
\left[ Z_m-\frac{1}{\sqrt3}\frac{1}{\sqrt{h(t)}}Z^{\prime}_m \right] ,
\label{e27}
\end{equation}
with $h(t)=1+4t^2$, for neutrinos and
\begin{equation}
{\cal L}_l^{NC} =-\frac{g}{4}\frac{M_Z}{M_W} \left[ \bar
l\gamma^m(v_l+a_l\gamma^5)lZ_m+ \bar
l\gamma^m(v'_l+a'_l\gamma^5)lZ^{\prime}_m \right] ,
\label{e28}
\end{equation}
for the charged leptons, where we have defined
\begin{eqnarray}
\begin{array}{clccc}
v_l= & -1/h(t),&a_l=& 1,& \nonumber \\
v'_l=& -\sqrt{3/h(t)},
& a'_l=& v'_l/3.& \nonumber
\end{array}
\nonumber
\end{eqnarray}
We can use muon collider to discover the new neutral $Z'^0$ boson using the reaction 
$\mu e \to \mu e$ it was shown at \cite{Montero:1999en,Montero:1998sv} that $A_{RL}(\mu e)$ asymmetry is
considerably enhanced.

The Lagrangian interaction among quarks
and the $Z^0$ is
\begin{eqnarray}
{\cal L}_{ZQ}=-\frac{g}{4}\frac{M_Z}{M_W}\sum_i \left[ \bar
\Psi_i\gamma^m(v^i+a^i\gamma^5)\Psi_i \right] Z_m,
\label{e31}
\end{eqnarray}
where $i=u,c,t,d,s,b,J_1,J_2,J_3$; with
\begin{eqnarray}
\begin{array}{rlrrc}
v^U=& (3+4t^2)/3h(t),&a^U=&-1,& \nonumber \\
v^D=&-(3+8t^2)/3h(t),& a^D=&1,& \nonumber \\
v^{J1}=&-20t^2/3h(t),& a^{J_1}=&0,& \nonumber \\
v^{J_2}=v^{J_3}= & 16t^2/3h(t),& a^{J_2}=a^{J_3}=&0,& \nonumber 
\end{array}
\nonumber
\end{eqnarray}
$U$, and $D$ mean the charge $+2/3$ and $-1/3$ respectively, the same
for $J_{1,2,3}$. There is also the usual QCD Lagrangian given by
\begin{eqnarray}
{\cal L}^{QCD}_{q}&=&=g_{s}G^{\mu}\bar{q}\gamma_{\mu}q \,\ ,
\label{lagr}
\end{eqnarray}
the lagrangians presented at Eqs.(\ref{lq},\ref{e27},\ref{e28},\ref{e31},\ref{lagr}) are the same as appear in the 
331 model \cite{ppf}. In those lagrangians appear a lot of interestings phenomenologiacal studies presented at \cite{Coutinho:1999hf}.

\begin{figure}[ht]
\begin{center}
\vglue -0.009cm
\mbox{\epsfig{file=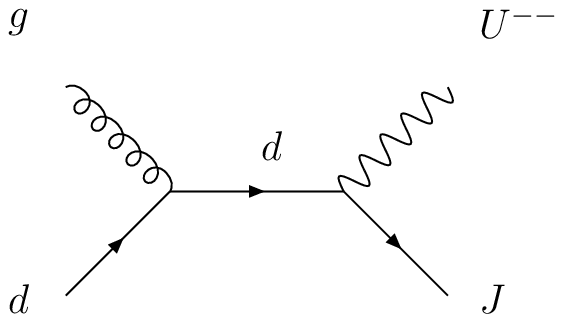,width=0.5\textwidth,angle=0}}       
\end{center}
\caption{$gd \to U^{--}J$ exchanging quark-$d$.}
\label{fig1a}
\end{figure}

\begin{figure}[ht]
\begin{center}
\vglue -0.009cm
\mbox{\epsfig{file=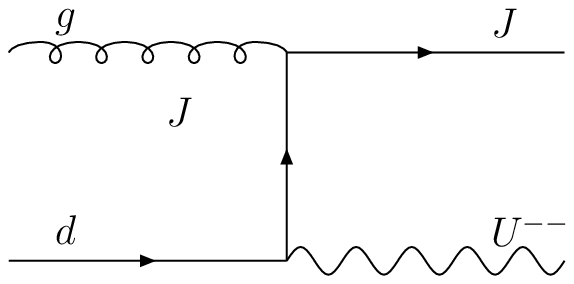,width=0.5\textwidth,angle=0}}
\end{center}
\caption{$gd \to U^{--}J$ exchanging quark-$J$.}
\label{fig1b}
\end{figure}

We can, also, study the following process resulting in at least three leptons coming from $pp$ collision, throught the following reactions 
\footnote{I would like to thanks to Alexander Belyaev that call my attention to the first process.}
\begin{eqnarray}
g+d &\to& U^{--}+J, \,\ g+u \to U^{++}+j_{\alpha}, \nonumber \\
d+ \bar{d}&\to&U^{++}U^{--}, \,\ d+ \bar{u} \to U^{--}+V^{+}.
\label{intersting}
\end{eqnarray}
For example the first process has the following Feynmann diagrams drawing in Figs.~(\ref{fig1a},\ref{fig1b}). Similar 
diagrams can be drawn to the process $g+u \to U^{++}+j_{\alpha}$ (change $d \rightarrow u$ and $J \rightarrow j$).

\begin{figure}[ht]
\begin{center}
\vglue -0.009cm
\mbox{\epsfig{file=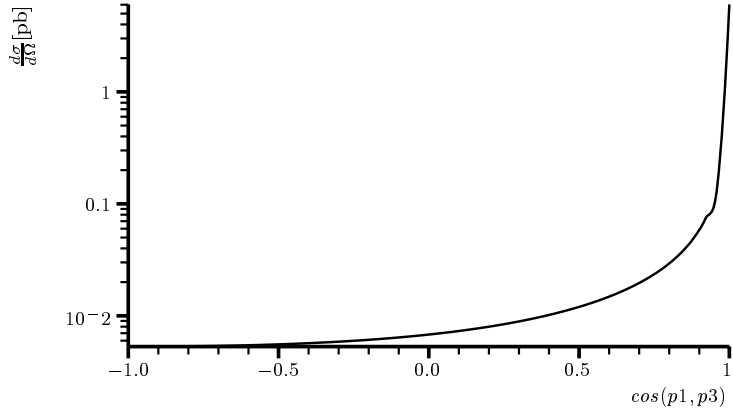,width=0.7\textwidth,angle=0}}       
\end{center}
\caption{Differential Cross Section $gd \to JU^{--}$.}
\label{choq1}
\end{figure}

\begin{figure}[ht]
\begin{center}
\vglue -0.009cm
\mbox{\epsfig{file=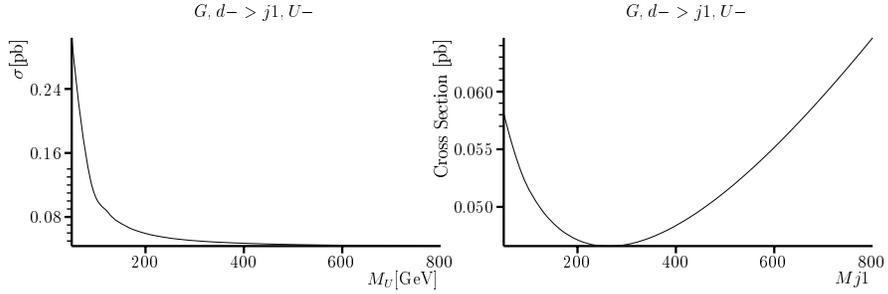,width=0.7\textwidth,angle=0}}       
\end{center}
\caption{Total Cross Section $gd \to JU^{--}$ as 
function of $M_U$~(left) and $M_{J}$~(right).}
\label{choq3}
\end{figure}

\begin{figure}[ht]
\begin{center}
\vglue -0.009cm
\mbox{\epsfig{file=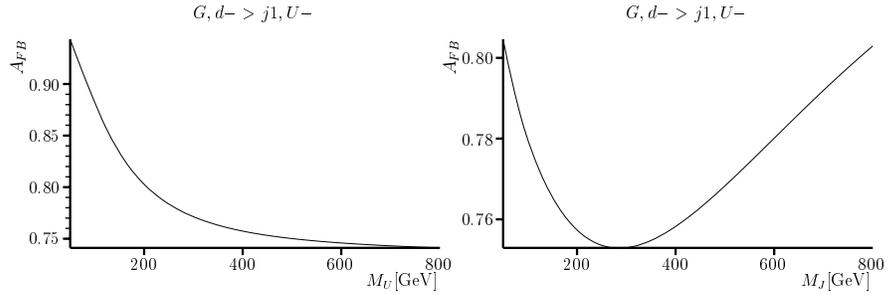,width=0.7\textwidth,angle=0}}       
\end{center}
\caption{Asymmetry $gd \to JU^{--}$ as function of
 $M_U$~(left) and $M_{J}$~(right).}
\label{assi2}
\end{figure}

As first results, in Fig.\ref{choq1} we present the differencial cross section, get from the program COMPHEP \cite{comphep}, to the process 
$gd \to JU^{--}$ as function of $cos(p1,p3)$. We have also calculated the cross section to 
these process as function of $M_{U}$ and $M_{J}$ and our results are shown in
Figs.\ref{choq3}~(left) and Figs.\ref{choq3}~(right) respectively. In Figs.~\ref{assi2}~left(right) we present the results 
on forward-baskward asymmetry as function of $M_{U}$($M_{J}$) respectivelly. Similar results can also be get to 
the process $g+u \to U^{++}+j_{\alpha}$.

\begin{figure}[ht]
\begin{center}
\vglue -0.009cm
\mbox{\epsfig{file=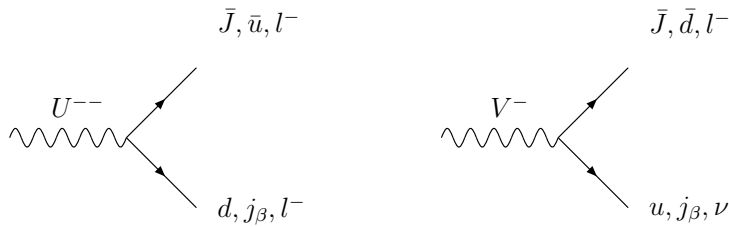,width=0.7\textwidth,angle=0}}       
\end{center}
\caption{U and V decay in two particles}
\label{fig3}
\end{figure}

From Eq.(\ref{lagr}), the new 
gauge bosons can decay in the followings channels 
$U^{--}\to ( \bar{J}d, \bar{u}j_{\beta},l^{-}l^{-})$ and 
$V^{-}\to ( \bar{J}u, \bar{d}j_{\beta},l^{-}\nu )$. These decays modes are 
shown in Fig.~\ref{fig3}.

\begin{figure}[ht]
\begin{center}
\vglue -0.009cm
\mbox{\epsfig{file=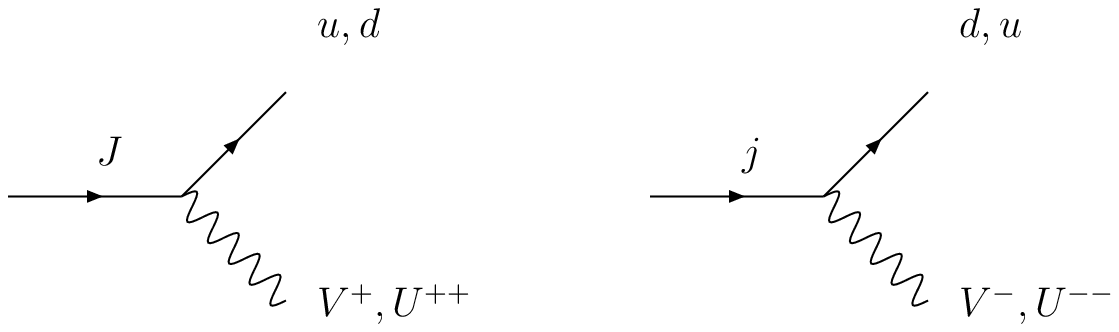,width=0.7\textwidth,angle=0}}       
\end{center}
\caption{J and j decay in ordinary quarks and bileptons that will decay too.}
\label{fig5}
\end{figure}

The heavy quarks $J$, $j_{1}$ and $j_{2}$ can decay to the light quark via 
$V^{*}/U^{*}$ emission to produce bilepton final states with a specific decay 
signature, see Fig.~\ref{fig5}. Analasing these decays mode, we conclude that the $J$ quark will decay 
in $l^{+}l^{+}d$ or $l^{+}\nu u$ without any restrictions coming 
from the bosons gauge masses, because these particles are virtual on this 
decay. While the $j$ quark can decay in $l^{-}\nu d$ and $l^{-}l^{-}\bar{u}$.

By another hand, the $U$ decay will depend of $M_{U}$, $M_{J}$ and 
$M_{j_{\beta}}$. in Tab.~\ref{t0} we shown all possibles possibilities.
\begin{table}
\begin{tabular}{|c|c|c|} 
\hline
Case number & Mass relation & decay mode   \\ 
\hline
1 & $M_{U}>M_{J}$, $M_{U}>M_{j}$&$\bar{J}d$, $\bar{u}d$, $l^{-}l^{-}$  \\ 
\hline
2 & $M_{U}<M_{J}$, $M_{U}>M_{j}$&$\bar{J}d$, $l^{-}l^{-}$  \\ \hline
3 & $M_{U}>M_{J}$, $M_{U}<M_{j}$&$\bar{u}d$, $l^{-}l^{-}$   \\ \hline
4 & $M_{U}<M_{J}$, $M_{U}<M_{j}$& $l^{-}l^{-}$  \\ \hline
\end{tabular}
\caption{All possibles decays to the $U$ boson.}
\label{t0}
\end{table}

\begin{figure}[ht]
\begin{center}
\vglue -0.009cm
\mbox{\epsfig{file=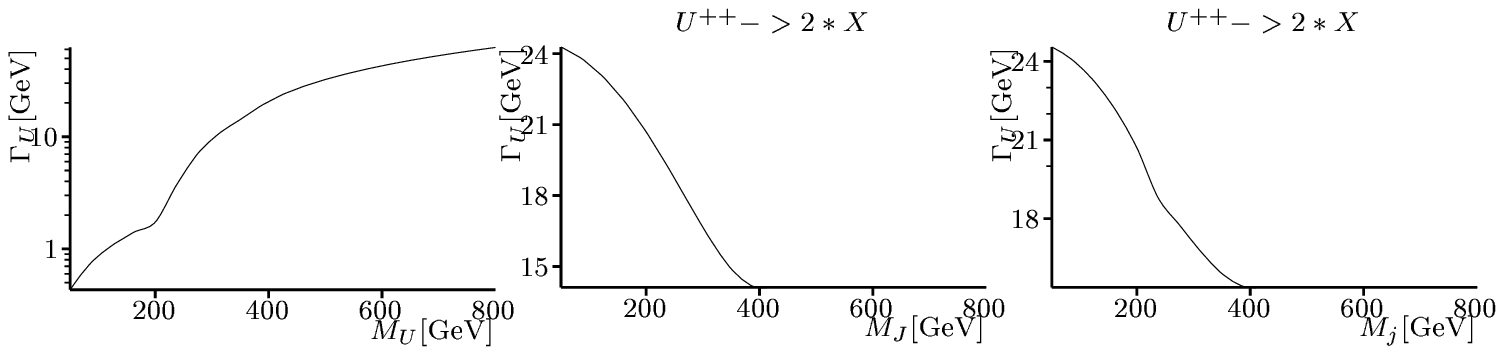,width=0.7\textwidth,angle=0}}       
\end{center}
\caption{\label{fig:u-width}
U decay width as function of $M_U$(left), $M_{J}$(center) and  $M_{j}$(right)}
\end{figure}

The width of the $U$ boson is drawing in the Fig.~\ref{fig:u-width}~(left) as 
function of its mass. In Fig.~\ref{fig:u-width}~(center)
we draw $\Gamma_{U}$ versus $M_{J}$, while in Fig.~\ref{fig:u-width}~(right)
we plot $\Gamma_{U}$ versus  $M_{j}$.

Again we divided the signals for the process $gd \to lllX$ in four regions.
\begin{table}
\begin{tabular}{|c|c|} 
\hline
$\underline{l^{+}l^{+}d}$ $\underbrace{l^{-}l^{-}\bar{d}d}$ & 
$\underline{l^{+}\bar{\nu}u}$ $\underbrace{l^{-}l^{-}\bar{d}d}$ \\ \hline
$\underline{l^{+}l^{+}d}$ $\underbrace{l^{-}\nu \bar{u}d}$ & 
$\underline{l^{+}\bar{\nu}u}$ $\underbrace{l^{-}\nu \bar{u}d}$ \\ \hline
$\underline{l^{+}l^{+}d}$ $\underbrace{l^{-}l^{-}\bar{u}u}$ & 
$\underline{l^{+}\bar{\nu}u}$ $\underbrace{l^{-}l^{-}\bar{u}u}$ \\ \hline
$\underline{l^{+}l^{+}d}$ $\underbrace{l^{-}l^{-}}$ & 
$\underline{l^{+}\bar{\nu}u}$ $\underbrace{l^{-}l^{-}}$ \\ \hline
\end{tabular}
\caption{States coming from $JU^{--}$ decay.}
\label{tab1}
\end{table}

\begin{figure}[ht]
\begin{center}
\vglue -0.009cm
\mbox{\epsfig{file=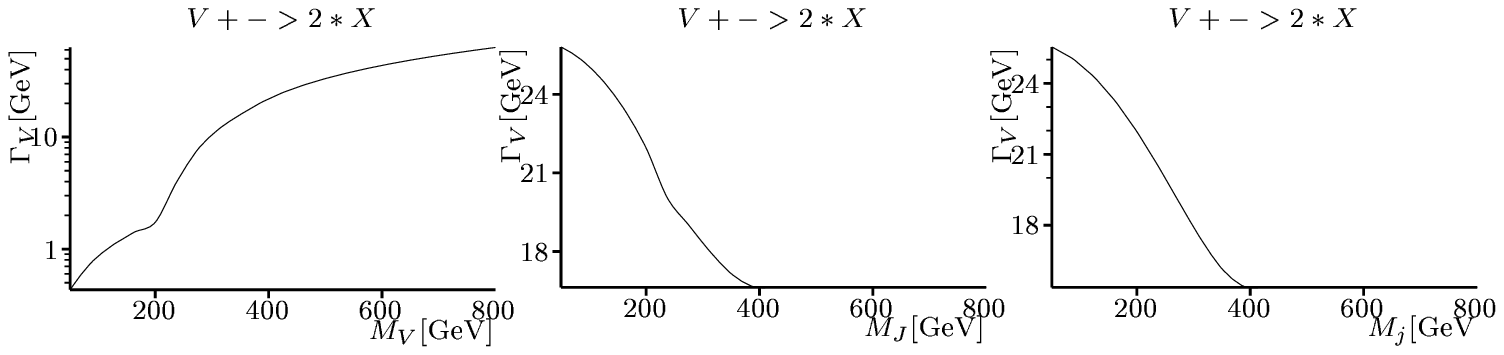,width=0.7\textwidth,angle=0}}       
\end{center}
\caption{\label{fig:v-width}
V decay width as function of $M_U$(left), $M_{J}$(center) and  $M_{j}$(right)}
\end{figure}

We also present the width of the $V$ boson
versus its mass in Fig.~\ref{fig:v-width}~(left) as function of 
its mass. In Fig.~\ref{fig:v-width}~(center)
$\Gamma_{V}$ versus $M_{J}$ is shown, while in Fig.~\ref{fig:u-width}~(right)
we plot $\Gamma_{V}$ versus  $M_{j}$.

Of course this process must be better studied as the others three processes listed at Eq.(\ref{intersting}). These 
particles can be detected at Large Hadron Collider (LHC) if they really exist in nature.

There are background come mainly from the SM and from MSSM \cite{dress,tata}. The background from the SM comes from the 
$W^{\star}Z^{\star}$, $W^{\star}\gamma^{\star}$,$Z^{\star}Z^{\star}$ and $\bar{q}q$. Where in the SM we 
have the following decays for the gauge bosons $W^{-} \to l^{-} \nu_{l}$, $Z^{0} \to l^{+}l^{-}$ and $\gamma \to l^{+}l^{-}$.

The $W^{\star}Z^{\star}$ and $W^{\star}\gamma^{\star}$ background are known to be the major source of background 
come from the SM for the three lepton channjel. The second largest background font to three leptons channel is 
from $q \bar{q}$ events. Finally, the remaining three leptons background which should worry about is the 
$Z^{\star}Z^{\star}$ jet production.

In the MSSM the charginos, neutralinos, gluinos and squarks pair production leads to a trilepton signature too. 
The trilepton final states that could arise from the decay of charginos $\tilde{\chi}^{\pm}_{1}$ and neutralinos 
$\tilde{\chi}_{2}$. For the reaction $\bar{q}q \to \tilde{\chi}^{\pm}_{1}\tilde{\chi}^{0}_{2}$, where 
$\tilde{\chi}^{\pm}_{1} \to \tilde{\chi}^{0}_{1}l^{\pm}\nu_{l}$ and 
$\tilde{\chi}^{0}_{2} \to \tilde{\chi}^{0}_{1}l^{+}l^{-}$, and $\tilde{\chi}^{0}_{1}$ is the LSP. The $\nu_{l}$ and 
two LSPs do not interact and manifest themselves as missing energy. The resulting final states is three isolated 
charged leptons plus missing energy. While the squarks $\tilde{q}$ and the gluinos $\tilde{g}$ have the following 
interactions $\bar{\tilde{q}}\tilde{q}$, $\tilde{q}\tilde{g}$ and $\tilde{g}\tilde{g}$ and the decays of squarks 
and gluinos are $\tilde{q} \to q\tilde{\chi}^{0}_{1}$ and $\tilde{g} \to \bar{q}q \tilde{\chi}^{0}_{1}$ 
\cite{Dawson,dress,tata}.

To finish this analyses from interactions of gauge bosons, we can study ${\cal L}_{dc}$ given at Eq.(\ref{tutty}) at 
Appendix \ref{sec:lagrangian}. From this lagrangian we can derive the following Feynman rules given the trilinear 
and quartic coupling. On this case we get the same results presented \cite{Hoang:2001km,Binh:2002an}, given 
at Tab(\ref{verticesautointeracao}).
\begin{table}[h]
\begin{center}
\begin{tabular}{|c|c|}  \hline \hline
Vertex & coupling constant/e   \\  \hline \hline
$\gamma W^+ W^-$  & 1\\  \hline
$Z W^+ W^-$ &$1/t_W$ \\ \hline
$\gamma V^+ V^-$&$1$\\ \hline
$Z V^+ V^-$ & $- (1 + 2 s_W^2)/\sin 2\theta_W$\\  \hline
$\gamma U^{++} U^{--}$&$ 2 $\\ \hline
$Z U^{++} U^{--}$&$(1 - 4 s_W^2)/\sin 2\theta_W$\\ \hline
$Z' V^+ V^-$&$ - \sqrt{3(1 - 4 s_W^2)}/\sin
2\theta_W$\\ \hline
$Z' U^{++} U^{--}$&$ -  \sqrt{3(1 - 4 s_W^2)}/\sin
2\theta_W$\\ \hline
$U^{--} V^+ W^+$&$ 1/(\sqrt{2}\  s_W)$\\ \hline
$U^{++} W^- V^-$&$ 1/(\sqrt{2} \ s_W)$\\ \hline
$ W^+_\mu W^-_\nu  W^+_\alpha W^-_\beta  $  & $ S_{\mu \alpha,\nu \beta}$\\ \hline 
$ V^+_\mu  V^-_\nu  V^+_\alpha  V^-_\beta $  &
$S_{\mu \alpha,\nu \beta}$ \\
\hline $ U^{++}_\mu  U^{--}_\nu  U^{++}_\alpha  U^{--}_\beta $  &
 $ S_{\mu \alpha,\nu \beta}$\\
\hline $ W^+_\mu  W^-_\nu V^+_\alpha   V^-_\beta $ & $ S_{\mu \beta,\nu \alpha }/2$\\
 \hline $ W^+_\mu  W^-_\nu  U^{++}_\alpha U^{--}_\beta$  &
  $  S_{\mu \alpha, \nu \beta}/2$\\
\hline $ V^{+}_\mu V^{-}_\nu U^{++}_\alpha U^{--}_\beta $ &
 $  S_{\mu \alpha, \nu \beta}/2 $\\
\hline $\gamma_\mu  \gamma_\nu  W^+_\alpha  W^-_\beta$ &
 $ - s^{2}_W S_{\mu \nu ,\alpha \beta} $\\
\hline $\gamma_\mu  \gamma_\nu  V^+_\alpha  V^-_\beta$ &
 $ - s^{2}_W  S_{\mu \nu ,\alpha \beta}$
\\  \hline $\gamma_\mu  \gamma_\nu  U^{++}_\alpha  U^{--}_\beta$  &
 $ - 4 s^{2}_W S_{\mu \nu ,\alpha \beta} $ \\
\hline $ Z_\mu  Z_\nu  W^+_\alpha  W^-_\beta $  &
 $ - c^{2}_W  S_{\mu \nu ,\alpha \beta} $\\
\hline $ Z_\mu  Z_\nu  V^{+}_\alpha V^{-}_\beta$ &
$ - (c_W - 3 s_W t_W)^2  S_{\mu \nu ,\alpha \beta}/4 $ \\
\hline $ Z_\mu  Z_\nu  U^{++}_\alpha  U^{--}_\beta $  &
 $ - (c_W - 3 s_W t_W)^2 S_{\mu \nu ,\alpha \beta}/4 $ \\
\hline $ Z'_\mu  Z'_\nu  V^{+}_\alpha V^{-}_\beta$ &
 $ - 3(1- 3t^2_W) S_{\mu \nu ,\alpha \beta}/4 $\\
\hline $ Z'_\mu  Z'_\nu  U^{++}_\alpha  U^{--}_\beta $  &
 $ - 3(1-3t^2_W)  S_{\mu \nu ,\alpha \beta}/4 $ \\
  \hline $ \gamma_\mu  Z_\nu  W^+_\alpha  W^-_\beta$  &
   $ - c_W s_W S_{\mu \nu ,\alpha \beta} $\\
\hline $ \gamma_\mu  Z_\nu  V^{+}_\alpha V^{-}_\beta$ &
 $  s_W(c_W + 3 s_W t_W) S_{\mu \nu ,\alpha \beta}/2  $\\
\hline $\gamma_\mu  Z_\nu  U^{++}_\alpha  U^{--}_\beta $  &
  $ -s_W(c_W - 3 s_W t_W) S_{\mu \nu ,\alpha \beta} $\\
\hline $ \gamma_\mu  Z'_\nu   V^{+}_\alpha  V^{-}_\beta$ &
 $  s_W\sqrt{(3 - 9 t_W^2)}  S_{\mu \nu ,\alpha \beta}/2 $\\
  \hline $\gamma_\mu  Z'_\nu  U^{++}_\alpha  U^{--}_\beta $  &
  $ s_W \sqrt{(3 - 9 t_W^2)} S_{\mu \nu ,\alpha \beta} $\\
  \hline $ Z_\mu   Z'_\nu  V^{+}_\alpha V^{-}_\beta$ &
  $ - (c_W + 3 s_W t_W)\sqrt{(3 - 9 t_W^2)} S_{\mu \nu ,\alpha \beta}/4$\\
  \hline $ Z_\mu   Z'_\nu  U^{++}_\alpha U^{--}_\beta $  &
   $(c_W - 3 s_W t_W) \sqrt{(3 - 9 t_W^2)} S_{\mu \nu ,\alpha \beta}/4$\\
\hline $ Z'_\mu  W^+_\nu  V^{+}_\alpha U^{--}_\beta$ &
 $  \sqrt{6(1-3t_W^2)} S_{\mu \nu ,\alpha \beta}/4$\\
\hline $\gamma_\mu  W^{+}_\nu V^{+}_\alpha U^{--}_\beta$ &
 $ 3 s_W  V_{\mu \nu \alpha \beta}/\sqrt{2}$\\
\hline $ Z_\mu   W^+_\nu  V^{+}_\alpha U^{--}_\beta$ &
 $ 3\left(s_W t_WS_{\mu \nu ,\alpha \beta }
 +c_W U_{\mu \beta \nu \alpha}\right)/(2 \sqrt{2}) $\\
\hline
\end{tabular}
\caption{Trilinear and Quartic couplings in the MSUSY331}
\end{center}
\label{verticesautointeracao}
\end{table}
Here the following notations were used
\begin{eqnarray}
 S_{\mu \nu,\alpha \beta}&\equiv&g_{\mu \alpha }g_{\nu \beta}+
 g_{\mu\beta}g_{\nu \alpha} -2g_{\mu \nu}g_{\alpha \beta}, \nonumber \\
 V_{\mu \nu \alpha \beta}& \equiv &g_{\mu \nu}g_{\alpha \beta}-
 g_{\mu \alpha}g_{\nu \beta},   \nonumber\\
 U_{\mu \beta \nu \alpha}&\equiv&  g_{\mu \beta}g_{ \nu \alpha}-
 g_{\mu \alpha}g_{\nu \beta}.
\end{eqnarray}

\section{$R$-Parity}
\label{sec:rparity}

The R-symmetry was introduced in 1975 by A. Salam and J. Strathdee
\cite{r1} and in an independent way by P. Fayet \cite{Fayet:1974pd} to avoid
the interactions that violate either lepton number or baryon
number. There is very nice review about this subject in
Refs.\cite{barbier,moreau}. More precisely, R-parity (which keeps
particles invariant, and changes the sign of  sparticles) can be
written as
\begin{equation}
R = (-1)^{3 (B - L) + 2 S}
\label{rparity}
\end{equation}
where $S$ is the spin of the particle.

We said above that only the total lepton number, $L$,
remains a global quantum number (or equivalently we can define ${\cal F}=
B+L$ as the global conserved quantum number where $B$ is the 
baryonic number~\cite{Pleitez:1992xh}). However, if we assume the global $U(1)_{\cal F}$ 
symmetry, it allows us to introduce the $R$-conserving symmetry,
defined as $R=(-1)^{3{\cal F}+2S}$.
The ${\cal F}$ number attribution is
\begin{equation}
\begin{array}{c}
{\cal F}(U^{--})={\cal F}(V^{-}) = - {\cal F}(J_1)= {\cal F}(J_{2,3})=
{\cal F}(\rho^{--})  \\  
= {\cal F}(\chi^{--}) ={\cal F}(\chi^{-}) = 
{\cal F}(\eta^-_2)=2,\end{array}
\label{efe}
\end{equation}
with ${\cal F}=0$ for the other Higgs scalar, while for leptons and the known 
quarks ${\cal F}$ coincides with the total lepton and baryon numbers, 
respectively.

Choosing the following R-charges 
\begin{eqnarray}
n_{\eta}&=&n_{\rho^{\prime}}=-1, \,\
n_{\rho}=n_{\eta^{\prime}}=1, \,\
n_{\chi}=n_{\chi^{\prime}}=0, \nonumber \\
n_{L}&=&n_{Q_{i}}=n_{d_{i}}=1/2, \,\
n_{J_{i}}=-1/2, \,\ n_{u}=-3/2,
\label{rdiscsusy331} 
\end{eqnarray} 
it is easy to see that all the fields $\eta$,
$\eta^{\prime}$, $\chi$, $\chi^{\prime}$, $\rho$, $\rho^{\prime}$,
$L$, $Q_{i}$, $u$, $d$ and $J_{i}$ have R-charge equal to one, while their superpartners
have opposite R-charge similar to that in the MSSM. 
The terms which satisfy the defined above symmetry (\ref{rdiscsusy331}) the 
term allowed by this $R$-charges in our superpotential, given at 
Eq.(\ref{sp3m1}), are given by
\begin{eqnarray} 
W&=&\mu_{ \eta} \hat{ \eta} \hat{ \eta}^{\prime}+
\mu_{ \rho} \hat{ \rho} \hat{ \rho}^{\prime}+
\mu_{ \chi} \hat{ \chi} \hat{ \chi}^{\prime}+
\lambda_{2ab} \epsilon \hat{L}_{aL} \hat{L}_{bL} \hat{ \eta}+
f_{1} \epsilon \hat{ \rho} \hat{ \chi} \hat{ \eta}+
f^{\prime}_{1}\epsilon \hat{ \rho}^{\prime}\hat{ \chi}^{\prime}\hat{ \eta}^{\prime} \nonumber \\ &+&
\kappa_{1\alpha i} \hat{Q}_{\alpha L} \hat{\rho} \hat{u}^{c}_{iL} +
\kappa_{2\alpha i} \hat{Q}_{\alpha L} \hat{\eta} \hat{d}^{c}_{iL} +
\kappa_{3\alpha \beta} \hat{Q}_{\alpha L} \hat{\chi} \hat{j}^{c}_{\beta L}+
\kappa_{4\alpha ai} \hat{Q}_{\alpha L} \hat{L}_{aL} \hat{d}^{c}_{iL}+
\kappa_{5i} \hat{Q}_{3L} \hat{\eta}^{\prime} \hat{u}^{c}_{iL}\nonumber \\ &+&
\kappa_{6i} \hat{Q}_{3L} \hat{\rho}^{\prime} \hat{d}^{c}_{iL}+
\kappa_{7} \hat{Q}_{3L} \hat{\chi}^{\prime} \hat{J}^{c}_{L}.
\end{eqnarray}
In this case only the quarks get masses. However not all of the leptons get mass. This is 
because the Yukawa coupling $\lambda_{2ab}$ is only non-zero when it is
antisymmetric in the generation indices ($a,b$). In the usual 331 model to generate the charged 
lepton masses we introduce an antisextet, as we don't introduce this scalar in our model the 
charged leptons are massless in this case. The neutrinos are also massless.

However, if we want to allow neutrinos to get their masses and at the same time avoid the 
fast nucleon decay we can choose the following $R$-charges
\begin{eqnarray}
n_{L}&=&n_{\eta}=n_{\eta^{\prime}}=n_{\rho}=n_{\rho^{\prime}}=-n_{\chi}=n_{\chi^{\prime}}=0, \nonumber \\
n_{Q_{i}}&=&1, \,\ n_{u}=n_{d_{i}}=n_{J_{i}}=-1.
\label{rsusy331nm} 
\end{eqnarray} 
In this case, the terms allow in our superpotential are
\begin{eqnarray} 
W&=&\mu_{0a}\hat{L}_{aL} \hat{ \eta}^{\prime}+\mu_{ \eta} \hat{ \eta} \hat{ \eta}^{\prime}+
\mu_{ \rho} \hat{ \rho} \hat{ \rho}^{\prime}+
\mu_{ \chi} \hat{ \chi} \hat{ \chi}^{\prime}+
\lambda_{1abc} \epsilon \hat{L}_{aL} \hat{L}_{bL} \hat{L}_{cL}+ 
\lambda_{2ab} \epsilon \hat{L}_{aL} \hat{L}_{bL} \hat{ \eta}\nonumber \\ &+&
\lambda_{3a} \epsilon \hat{L}_{aL} \hat{\chi} \hat{\rho}+
f_{1} \epsilon \hat{ \rho} \hat{ \chi} \hat{ \eta} +
f^{\prime}_{1}\epsilon \hat{ \rho}^{\prime}\hat{ \chi}^{\prime}\hat{ \eta}^{\prime}+
\kappa_{1\alpha i} \hat{Q}_{\alpha L} \hat{\rho} \hat{u}^{c}_{iL} +
\kappa_{2\alpha i} \hat{Q}_{\alpha L} \hat{\eta} \hat{d}^{c}_{iL} \nonumber \\ &+&
\kappa_{3\alpha \beta} \hat{Q}_{\alpha L} \hat{\chi} \hat{j}^{c}_{\beta L}+
\kappa_{4\alpha ai} \hat{Q}_{\alpha L} \hat{L}_{aL} \hat{d}^{c}_{iL}+
\kappa_{5i} \hat{Q}_{3L} \hat{\eta}^{\prime} \hat{u}^{c}_{iL} +
\kappa_{6i} \hat{Q}_{3L} \hat{\rho}^{\prime} \hat{d}^{c}_{iL}\nonumber \\ &+&
\kappa_{7} \hat{Q}_{3L} \hat{\chi}^{\prime} \hat{J}^{c}_{L}.
\label{suppotrsusy331nm}
\end{eqnarray}
In our superpotential, we can generate mass to neutrinos, as we will show in the next section, 
and we get that the nucleon is stable at tree-level \cite{longpal}. 
However it is not enough to forbid
the dangerous processes of nucleon decay but also forbid the neutron-antineutron oscillation, 
see Refs.~\cite{dress,tata,barbier,moreau,drei}. 

The last term in this superpotential induce the following nice process\cite{dress,tata,barbier}
\begin{enumerate}
\item Double Beta Decay without Neutrinos
\item New contributions to the Neutrals $K\bar{K}$ and also
$B\bar{B}$ Systems;
\item An additional contribution to the muon decay;
\item Charged Current Universality in $\pi$ and $\tau$ decays;
\item Charged Current Universality in the Quark Sector;
\item Leptonic Decays of Heavy Quarks Hadrons such as 
$D^{+} \rightarrow \overline{K^{0}}l^{+}_{i}\nu_{i}$;
\item Rare Leptonic Decays of Mesons like $K^{+} \rightarrow \pi^{+}\nu \bar{\nu}$,
\item Hadronic $B$ Meson Decay Asymmetries.
\end{enumerate}
it also give the following direct decays of the
lightest neutralinos
\begin{eqnarray}
\tilde{\chi}^{0}_{1} &\rightarrow& l^{+}_{i}\bar{u}_{j}d_{k}, \,\
\tilde{\chi}^{0}_{1} \rightarrow l^{-}_{i}u_{j}\bar{d}_{k},
\nonumber \\
\tilde{\chi}^{0}_{1} &\rightarrow& \bar{\nu}_{i}\bar{d}_{j}d_{k},
\,\ \tilde{\chi}^{0}_{1} \rightarrow \nu_{i}d_{j}\bar{d}_{k},
\end{eqnarray}
and for lightest charginos
\begin{eqnarray}
\tilde{\chi}^{+}_{1} &\rightarrow& l^{+}_{i}\bar{d}_{j}d_{k}, \,\
\tilde{\chi}^{+}_{1} \rightarrow l^{+}_{i}\bar{u}_{j}u_{k},
\nonumber \\
\tilde{\chi}^{+}_{1} &\rightarrow& \bar{\nu}_{i}\bar{d}_{j}u_{k},
\,\ \tilde{\chi}^{+}_{1} \rightarrow \nu_{i}u_{j}\bar{d}_{k}.
\end{eqnarray}
These very nice aspects also happen in SUSY331RN and in SUSYECO331 models \cite{Long}.

\section{Chargino and Neutralino Production}
\label{sec:feno2}

However, on this model we have doubly charged vector bosons and scalars, respectivelly. This means 
that in some supersymmetric extensions of these kind of models we will have double charged charginos 
\cite{mcr,Rodriguez:2007yy,Rodriguez:2007jc}. On this model the charginos can decay in 
the following way
\begin{eqnarray}
\tilde{\chi}^{++}&\rightarrow& \tilde{l}^{+}l^{+}, \nonumber \\
\tilde{\chi}^{+}&\rightarrow& \tilde{\nu}l^{+}, \nonumber \\
\tilde{\chi}^{0}&\rightarrow& \tilde{\nu}\nu .
\label{ilc1}
\end{eqnarray}
Take this information into account we can say  that
\begin{eqnarray}
e^{-}e^ {-}&\rightarrow& \tilde{\chi}^{++}\tilde{\chi}^{0} 
\rightarrow \mbox{leptons} + \mbox{mixing energy}.
\end{eqnarray}
Again another interesting signal that can be measured at the International Linear Collider (ILC).

In a previous work \cite{mcr,Rodriguez:2007yy,Rodriguez:2007jc} we have have calculated the total 
cross section to the reactions
\begin{eqnarray}
e^{-}e^{-} &\to& \tilde{ \chi}^{-} \tilde{ \chi}^{-}, \nonumber \\
e^{-}e^{-} &\to& \tilde{ \chi}^{--} \tilde{ \chi}^{0},
\label{ilc2}
\end{eqnarray}
we know that the ILC will start to run with $\sqrt{s}=0,5$TeV and therefore this detector can detect 
them if they really exist in nature.

Naturally these particles can also be detected at LHC through the processes
\begin{eqnarray}
pp &\to& \tilde{ \chi}^{+} \tilde{ \chi}^{+}, \nonumber \\
pp &\to& \tilde{ \chi}^{++} \tilde{ \chi}^{0},
\label{ilc3}
\end{eqnarray}
and on this case we thinks it will be intersting to study these process to the LHC.

\section{Mass spectrum}
\label{sec:massspectrum}

Here in this article, we want to present the mass spectrum of the Minimal 
Supersymmetric 3-3-1 model. We will present first the results in the fermion's sector, then 
in the boson's sector. 

\subsection{Lepton´s masses}

In a previous work we have shown that, in the MSUSY331, 
we don't need to use the antisextet to generate the masses to the leptons.

Let us first considered the charged lepton masses.
Denoting
\begin{equation}
\begin{array}{c}
\phi^+=(e^c,\mu^c,\tau^c,-i\lambda^+_W,-i\lambda^+_V, 
\tilde{\eta}^{\prime+}_1,\tilde{\eta}^+_2,
\tilde{\rho}^+,\tilde{\chi}^{\prime+})^T,\\
\phi^-=(e,\mu,\tau,-i\lambda^-_W,-i\lambda^-_V, 
\tilde{\eta}^{-}_1,\tilde{\eta}^{\prime -}_2,
\tilde{\rho}^{\prime -},\tilde{\chi}^{-})^T,
\end{array}
\label{cbasis}
\end{equation}
where all the fermionic fields are still Weyl spinors, we can also, as before, 
define $\Psi^{\pm}=(\phi^+ \phi^-)^T$, and the mass term 
$-(1/2)[  
\Psi^{\pm T}Y^\pm\Psi^{\pm}+H.c.]$
where $Y^\pm$ is given by: 
\begin{equation}
Y^{\pm}= \left( \begin{array}{cc}
0  & X^T \\
X  & 0
\end{array}
\right),
\label{ypm}
\end{equation}
with
\begin{equation}
X= \left( \begin{array}{ccccccccc}
0&- \frac{\lambda_{2e \mu}}{3}v&- \frac{ \lambda_{2e \tau}}{3}v& 0& 0&- 
\frac{\mu_{0e}}{2}& 0&- \frac{\lambda_{3e}}{3}w& 0\\
\frac{ \lambda_{2e \mu}}{3}v& 0&- \frac{ \lambda_{2\mu \tau}}{3}v& 0& 0&- 
\frac{\mu_{0\mu}}{2}& 0&- \frac{\lambda_{3 \mu}}{3}w& 0\\
\frac{ \lambda_{2e \tau}}{3}v& \frac{ \lambda_{2\mu \tau}}{3}v& 0& 0& 0&- 
\frac{\mu_{0\tau}}{2}& 0&- \frac{\lambda_{3 \tau}}{3}w& 0\\
0& 0& 0& m_{\lambda}& 0&- gv^{\prime}& 0& gu& 0\\
0& 0& 0& 0& m_{\lambda}& 0& gv& 0&- gw^{\prime}\\
0& 0& 0& gv& 0&- \frac{\mu_{\eta}}{2}& 0& \frac{f_{1}w}{3}& 0\\
- \frac{\mu_{0e}}{2}&- \frac{\mu_{0\mu}}{2}&- \frac{\mu_{0\tau}}{2}& 0&- 
gv^{\prime}& 0&- \frac{\mu_{\eta}}{2}& 0&- \frac{f^{\prime}_{1}u^{\prime}}{3}\\
0& 0& 0&- gu^{\prime}& 0& \frac{f^{\prime}_{1}w^{\prime}}{3}& 0&- 
\frac{\mu_{\rho}}{2}& 0\\
- \frac{\lambda_{3e}}{3}u&- \frac{\lambda_{3 \mu}}{3}u&- 
\frac{\lambda_{3 \tau}}{3}u& 0& gw& 0&- \frac{f_{1}u}{3}& 0&- 
\frac{\mu_{\chi}}{2}
\end{array}
\right),
\label{clmm}
\end{equation}
where we have defined $v,v^{\prime},u,u^{\prime},w$ and $w^{\prime}$ at Eq.(\ref{vev1}).

The chargino mass matrix $Y^\pm$ is diagonalized using two unitary matrices, $D$
and $E$, defined by
\begin{eqnarray}
\tilde{ \chi}^{+}_{i}=D_{ij} \Psi^{+}_{j}, \,\
\tilde{ \chi}^{-}_{i}=E_{ij} \Psi^{-}_{j}, \,\ i,j=1, \cdots , 9,
\label{2sc} 
\end{eqnarray}
($D$ and $E$ sometimes are denoted, in non-supersymmetric theories, by $U^l_R$
and $U^l_L$, respectively). Then we can write the diagonal mass matrix as 
\begin{equation}
M_{SCM}=E^{*}XD^{-1}.
\label{m1}
\end{equation}
To determine $E$ and $D$, we note that
\begin{equation}
M^{2}_{SCM}=DX^T \cdot XD^{-1}=E^{*}X \cdot X^T(E^{*})^{-1},
\label{m2}
\end{equation}
and define the following Dirac spinors:
\begin{eqnarray}
\Psi(\tilde{ \chi}^{+}_{i})= \left( \begin{array}{cc}
             \tilde{ \chi}^{+}_{i} &
	     \bar{ \tilde{\chi}}^{-}_{i} 
\end{array} \right)^T, \,\
\Psi^{c}(\tilde{ \chi}^{-}_{i})= \left( \begin{array}{cc}
             \tilde{ \chi}^{-}_{i} &
	     \bar{ \tilde{\chi}}^{+}_{i} 
\end{array} \right)^T,
\label{emasssim}
\end{eqnarray}
where $\tilde{ \chi}^{+}_{i}$ is the particle and $\tilde{ \chi}^{-}_{i}$ 
is the anti-particle.

We have obtained the following masses (in GeV) for the charged sector:
\begin{eqnarray}
m_{\tilde{\chi}^{\pm}_{9}}&=&3186.05, \,\ m_{\tilde{\chi}^{\pm}_{8}}=3001.12, \,\ m_{\tilde{\chi}^{\pm}_{7}}=584.85, \nonumber \\ 
m_{\tilde{\chi}^{\pm}_{6}}&=&282.30, \,\ m_{\tilde{\chi}^{\pm}_{5}}= 204.55, \,\ m_{\tilde{\chi}^{\pm}_{4}}=149.41, 
\label{cm}
\end{eqnarray}
and the masses for the usual leptons (in GeV) $m_e=0$, $m_\mu=0.1052$ and 
$m_\tau=1.777$. 

From the first processes given at Eqs.(\ref{ilc2},\ref{ilc3}) we can detect at ILC or LHC 
several charginos (since $\tilde{\chi}^{\pm}_{1}$ until at least $\tilde{\chi}^{\pm}_{7}$) of this model.

These values have been obtained by
using  the following values for the dimensionless parameters
\begin{eqnarray}
\lambda_{2e\mu}=0.001,\;\lambda_{2e\tau}=0.001,\;\lambda_{2\mu\tau}=0.393,
\nonumber \\
\lambda_{3e}=0.0001,\; \lambda_{3\mu}=1.0,\;\lambda_{3\tau}=1.0,
\label{lambdas}
\end{eqnarray}
\begin{equation}
f_{1}=0.254,\;f^{\prime}_{1}=1.0,
\label{efes}
\end{equation}
and for the mass dimension parameters (in GeV) we have used:
\begin{equation}
\mu_{0e}=\mu_{0\mu}=0.0,\quad \mu_{0\tau}=10^{-6},
\label{mus0}
\end{equation}
\begin{equation}
\mu_{ \eta}=300,\; \mu_{ \rho}=500, \;\mu_{ \chi}=700,\;  m_{\lambda}=3000. 
\label{para}
\end{equation}
We also use the constraint $V^{2}_{ \eta}+V^{2}_{ \rho}=(246\;{\rm GeV})^2$ 
coming from $M_{W}$, where, we have defined 
$V^{2}_{ \eta}=v^{2}_{ \eta}+v^{ \prime 2}_{ \eta}$ and 
$V^{2}_{ \rho}=v^{2}_{ \rho}+v^{ \prime 2}_{ \rho}$.
Assuming that 
\begin{equation}
v_{ \eta}=20 GeV, \,\ v^{ \prime}_{ \eta}=v^{ \prime}_{ \rho}=1 {\mbox GeV}, {\mbox and} \,\ 
2v_{ \chi}=v^{ \prime}_{ \chi}=2 {\mbox TeV},
\label{valoresvev}
\end{equation}
 the
value of $v_\rho$ is fixed by the constraint above. 

Notice, from Eq.~(\ref{cm}), that the electron is massless at the tree level. 
This is a result of the structure of the mass matrix in 
Eq.~(\ref{clmm}) and there is not a symmetry that protects the electron 
to get a mass by loop corrections. We obtain that the dominant 
contribution to the electron mass is, up to logarithmic corrections, 
\begin{equation}
m_e\propto \lambda^\prime_{\alpha ei}\lambda^\prime_{\alpha' ej} 
V^2_j V^2_b
(v^2_\chi+v^2_{\chi'})\,\frac{m_{j_\alpha}}{9m^2_{\tilde{b}}},
\label{emass}
\end{equation}
and with all the indices fixed, $V_j$ denotes mixing matrix elements in the two
dimension $j_{1,2}$ space, $V_b$ means the same but in the d-like squark sector. 
We obtain $m_e=0.0005$ GeV if $v_\chi$ and $v^\prime_\chi$ have the values
already giving above.

\subsection{Neutralinos}

Like in the case of the charged sector, the neutral lepton masses are given 
by the mixing among neutrinos, gauginos and higgsinos. The mass term in the basis
\begin{equation}
\Psi^{0}= \left( 
\nu_{e} \nu_{\mu} \nu_{\tau}
-i \lambda^{3}_{A}
-i \lambda^{8}_{A}
-i \lambda_{B}
\tilde{\eta}^{0}
\tilde{\eta}^{ \prime 0}
\tilde{\rho}^{0}
\tilde{\rho}^{ \prime 0}
\tilde{\chi}^{0}
\tilde{\chi}^{ \prime 0}
\right)^T, 
\end{equation}
is given by $-(1/2)[ \left( \Psi^{0} 
\right)^T Y^{0} \Psi^{0}+H.c.] $ where
{\tiny
\begin{equation}
Y^{0}= \left( \begin{array}{cccccccccccc}
0& 0& 0& 0& 0& 0& 0&- \frac{\mu_{0e}}{2}& \frac{ \lambda_{3 e}}{3}w& 0& 
\frac{ \lambda_{3e}}{3}u& 0\\
0& 0& 0& 0& 0& 0& 0&- \frac{\mu_{0\mu}}{2}& \frac{ \lambda_{3 \mu}}{3}w& 0& 
\frac{ \lambda_{3\mu}}{3}u& 0\\
0& 0& 0& 0& 0& 0& 0&- \frac{\mu_{0\tau}}{2}& \frac{ \lambda_{3 \tau}}{3}w& 0& 
\frac{ \lambda_{3 \tau}}{3}u& 0\\
0& 0& 0& m_{\lambda}& 0& 0& \frac{gv}{\sqrt{2}}&- 
\frac{gv^{\prime}}{\sqrt{2}}&- 
\frac{gu}{\sqrt{2}}& \frac{gu^{\prime}}{\sqrt{2}}& 0& 0\\
0& 0& 0& 0& m_{\lambda}& 0& \frac{gv}{\sqrt{6}}&- \frac{gv^{\prime}}{\sqrt{6}}& 
\frac{gu}{\sqrt{6}}&- \frac{gu^{\prime}}{\sqrt{6}}&- \frac{2}{\sqrt{6}} gw& 
\frac{2}{\sqrt{6}}gw^{\prime}\\
0& 0& 0& 0& 0& m^{\prime}& 0& 0& \frac{g^{\prime}u}{\sqrt{2}}&- 
\frac{g^{\prime}u^{\prime}}{\sqrt{2}}&-  \frac{g^{\prime}w}{\sqrt{2}}& 
\frac{g^{\prime}w^{\prime}}{\sqrt{2}}\\
0& 0& 0&  \frac{gv}{\sqrt{2}}& \frac{gv}{\sqrt{6}}& 0& 0&- \frac{\mu_{\eta}}{2}&- 
\frac{f_{1}w}{3}& 0& \frac{f_{1}u}{3}& 0\\
- \frac{\mu_{0e}}{2}&- \frac{\mu_{0\mu}}{2}&- \frac{\mu_{0\tau}}{2}&-
\frac{gv^{\prime}}{\sqrt{2}}&-  
\frac{gv^{\prime}}{\sqrt{6}}& 0&- \frac{\mu_{\eta}}{2}& 0& 0&- 
\frac{f^{\prime}_{1}w^{\prime}}{3}& 0& \frac{f^{\prime}_{1}u^{\prime}}{3}\\
\frac{ \lambda_{3e}}{3}w& \frac{ \lambda_{3 \mu}}{3}w& 
\frac{ \lambda_{3 \tau}}{3}w&- 
\frac{gu}{\sqrt{2}}& \frac{gu}{\sqrt{6}}& \frac{g^{\prime}u}{\sqrt{2}}&- 
\frac{f_{1}w}{3}& 0& 0&- \frac{ \mu_{\rho}}{2}&- \frac{f_{1}v}{3}& 0\\
0& 0& 0&  \frac{gu^{\prime}}{\sqrt{2}}&- \frac{gu^{\prime}}{\sqrt{6}}&- 
\frac{g^{\prime}u^{\prime}}{\sqrt{2}}& 0&- \frac{f^{\prime}_{1}w^{\prime}}{3}&- 
\frac{ \mu_{\rho}}{2}& 0& 0&- \frac{f^{\prime}_{1}v^{\prime}}{3}\\
\frac{ \lambda_{3e}}{3}u& \frac{ \lambda_{3\mu}}{3}u& 
\frac{ \lambda_{3 \tau}}{3}u& 0&
- \frac{2}{\sqrt{6}} gw&- \frac{g^{\prime}w}{\sqrt{2}}& \frac{f_{1}u}{3}& 0&- 
\frac{f_{1}v}{3}& 0& 0&- \frac{ \mu_{\chi}}{2}\\
0& 0& 0& 0& \frac{2}{\sqrt{6}} gw^{\prime}& \frac{g^{\prime}w^{\prime}}{\sqrt{2}}& 0& 
\frac{f^{\prime}_{1}u^{\prime}}{3}& 0&- \frac{f^{\prime}_{1}v^{\prime}}{3}&- 
\frac{ \mu_{\chi}}{2}& 0
\end{array}
\right).
\label{mmn}
\end{equation}
}
All parameters in Eq.~(\ref{mmn}), but $m^\prime$, are defined in
Eqs.~(\ref{vev1}), (\ref{lambdas}) and (\ref{para}); $g$ and $g^\prime$ denote
the gauge coupling constant of $SU(3)_L$ and $U(1)_N$, respectively. 

The neutralino mass matrix is diagonalized by a $12 \times 12$ rotation unitary 
matrix $N$, satisfying
\begin{equation}
M_{NMD}=N^{*}Y^{0}N^{-1},
\end{equation}
and the mass eigenstates are
\begin{eqnarray}
\tilde{ \chi}^{0}_{i}&=&N_{ij} \Psi^{0}_{j}, \,\ j=1, \cdots ,12.
\label{emasneu}
\end{eqnarray}

We can define the following Majorana spinor to represent the mass eigenstates
\begin{eqnarray}
\Psi(\tilde{ \chi}^{0}_{i})&=& \left( \begin{array}{cc}
             \tilde{ \chi}^{0}_{i} &
	     \bar{ \tilde{\chi}}^{0}_{i} 
\end{array} \right)^T. 
\label{emassneu} 
\end{eqnarray}
As above the subindices $a,b,c$ run over the lepton generations $e,\mu,\tau$.

With the mass matrix in Eq.~(\ref{mmn}), at the tree level we obtain the
eigenvalues (in GeV), 
\begin{equation}
\begin{array}{c}
m_{\tilde{\chi}^{0}_{12}}=-4162.22, \,\ m_{\tilde{\chi}^{0}_{11}}=3260.48, \,\ m_{\tilde{\chi}^{0}_{10}}=3001.11, \\ 
m_{\tilde{\chi}^{0}_{9}}=585.19, \,\ m_{\tilde{\chi}^{0}_{8}}=-585.19, \,\ m_{\tilde{\chi}^{0}_{7}}=453.22,\\ 
m_{\tilde{\chi}^{0}_{6}}=-344.14, \,\ m_{\tilde{\chi}^{0}_{5}}=283.14, \,\ m_{\tilde{\chi}^{0}_{4}}=-272.0,
\end{array}
\label{hnm}
\end{equation}
and for the three neutrinos we obtain (in eV)
\begin{equation}
m_{\tilde{\chi}^{0}_{1}}=0,\;m_{\tilde{\chi}^{0}_{2}}\approx -0.01,\; m_{\tilde{\chi}^{0}_{3}}\approx 1.44.
\label{mnus1}
\end{equation}
We have got the values in Eqs.~(\ref{hnm}) and (\ref{mnus1}) by  choosing,
besides the parameters in Eqs.~(\ref{lambdas}) and (\ref{para}), 
$m^\prime=-3780.4159$ GeV. Notice that the coupling constant $g^\prime$ and the parameter $m^\prime$ 
appear only in the mass matrix of the neutralinos, all the other parameters in 
Eq.~(\ref{mmn}) have already been fixed by the charged sector, 
see Eq.~(\ref{clmm}), (\ref{lambdas}) and (\ref{para}). The neutrino masses in
Eq.~(\ref{mnus1}) are of the order of magnitude for LSND and solar neutrino
data.

\subsection{Double Charged Charginos}

Introducing the notation
\begin{eqnarray}
\psi^{++}= \left( \begin{array}{rrrrr}
-i \lambda^{++}_{U}&
\tilde{\rho}^{++}&
\tilde{\chi}^{ \prime ++}
\end{array}
\right)^t, \,\
\psi^{--}= \left( \begin{array}{rrrrr}
-i \lambda^{--}_{U}&
\tilde{\rho}^{ \prime --}&
\tilde{\chi}^{--}
\end{array}
\right)^t, \nonumber
\end{eqnarray}
and
\begin{equation}
\Psi^{\pm \pm}= \left( \begin{array}{rr}
\psi^{++} &
\psi^{--}
\end{array}
\right)^{t},
\end{equation}
we can write the following equation \cite{mcr}
\begin{eqnarray}
{\cal L}^{\mbox{double}}_{\mbox{mass}}&=&- \frac{1}{2} \left( \Psi^{\pm \pm} 
\right)^{t} Y^{\pm \pm} \Psi^{\pm \pm}+hc,
\end{eqnarray}
where
\begin{equation}
Y^{\pm \pm}= \left( \begin{array}{cc}
0  & T^{t} \\
T  & 0
\end{array}
\right),
\label{y++}
\end{equation}
with
\begin{equation}
T= \left( \begin{array}{ccccc}
-m_{\lambda}& -gu& gw^{\prime}& \frac{gz}{\sqrt{2}}& - \frac{gz^{\prime}}{\sqrt{2}} \\
gu^{\prime}& \frac{\mu_{\rho}}{2}&- 
\left( \frac{f_{1}^{\prime}v^{\prime}}{3}- \sqrt{2} \frac{f_{3}^{\prime}}{3}z^{\prime} \right)& 
0& \frac{f_{3}^{\prime}}{3}w^{\prime} \\
-gw&- \left( \frac{f_{1}v}{3}- \sqrt{2} \frac{f_{3}}{3}z \right)& \frac{\mu_{\chi}}{2}& 
\frac{f_{3}}{3}u& 0 \\
- \frac{gz^{\prime}}{\sqrt{2}}& 0& \frac{f_{3}^{\prime}}{3}u^{\prime}& \frac{\mu_{S}}{2}& 0 \\
\frac{gz}{\sqrt{2}}& \frac{f_{3}}{3}w& 0& 0& \frac{\mu_{S}}{2}
\end{array}
\right).
\label{MSSM1}
\end{equation}

The matrix $Y^{\pm \pm}$ in Eq.(\ref{y++}) satisfy the 
following relation
\begin{eqnarray}
\det (Y^{\pm \pm}- \lambda I)= \det \left[ \left( \begin{array}{cc}
- \lambda & T^{t} \\
T  &- \lambda 
\end{array} \right) \right]= \det( \lambda^{2}-T^{t} \cdot T),
\label{propmat1}
\end{eqnarray}
so we only have to calculate $T^{t} \cdot T$ to obtain the eigenvalues. 
Since $T^{t} \cdot T$ is a symmetric matrix, $\lambda^2$ must be real, and 
positive because $Y^{\pm \pm}$ is also symmetric. 

The double chargino mass matrix is diagonalized using two rotation matrices, 
$A$ and $B$, defined by
\begin{eqnarray}
\tilde{ \chi}^{++}_{i}=A_{ij} \Psi^{++}_{j}, \,\
\tilde{ \chi}^{--}_{i}=B_{ij} \Psi^{--}_{j}, \,\ i,j=1, \cdots , 5. 
\label{2dc}
\end{eqnarray}
where $A$ and $B$ are unitary matrices such that
\begin{equation}
M_{DCC}=B^{*}TA^{-1},
\end{equation}
the matrix $T$ is defined in Eq.(\ref{MSSM1}). To determine $A$ and $B$, we 
note that
\begin{equation}
M^{2}_{DCC}=AT^{t} \cdot TA^{-1}=B^{*}T \cdot T^{t}(B^{*})^{-1},
\end{equation}
which means that $A$ diagonalizes $T^{t} \cdot T$ while $B$ diagonalizes 
$T \cdot T^{t}$. It means
\begin{equation}
\mbox{diag}(m_{\tilde{\chi}^{\pm \pm}}) \equiv [B^{*}TA^{-1}]_{ij}=
m_{\tilde{\chi}^{\pm \pm}_{i}}\delta_{ij}.
\end{equation}

Performing the diagonalization we get the following numerical results in $GeV$
\begin{eqnarray}
m_{\tilde{\chi}^{\pm \pm}_{1}}&=&194.4, \,\
m_{\tilde{\chi}^{\pm \pm}_{2}}=343.3, \,\
m_{\tilde{\chi}^{\pm \pm}_{3}}=452.2, \,\
m_{\tilde{\chi}^{\pm \pm}_{4}}=652.1, \,\
m_{\tilde{\chi}^{\pm \pm}_{5}}=3187. \nonumber
\end{eqnarray}

Using this equation togheter Eqs(\ref{hnm},\ref{mnus1}) and considerating the second processes given at 
Eqs.(\ref{ilc2},\ref{ilc3}) we can detect at ILC or LHC 
several double charginos (since $\tilde{\chi}^{\pm \pm}_{1}$ until at least $\tilde{\chi}^{\pm \pm}_{4}$) and
neutralinos (since $\tilde{\chi}^{0}_{1}$ until at least $\tilde{\chi}^{0}_{9}$) of this model.

We define the following Dirac spinors to represent the mass 
eigenstates:
\begin{eqnarray}
\Psi(\tilde{ \chi}^{++}_{i})= \left( \begin{array}{cc}
             \tilde{ \chi}^{++}_{i} &
	     \bar{ \tilde{\chi}}^{--}_{i} 
\end{array} \right)^{t}, \,\
\Psi^{c}(\tilde{ \chi}^{--}_{i})= \left( \begin{array}{cc}
             \tilde{ \chi}^{--}_{i} &
	     \bar{ \tilde{\chi}}^{++}_{i} 
\end{array} \right)^{t},
\label{emassdou}
\end{eqnarray}
where $\tilde{ \chi}^{++}_{i}$ is the particle and $\tilde{ \chi}^{--}_{i}$ 
is the anti-particle, (we are using the same notation as in \cite{haber}). 

\subsection{Quark´s masses}

Let us first considered the u-quarks type.There are interactions like
\begin{eqnarray}
- \left[ \frac{ \kappa _{1i}}{3} \left( Q_{3} \eta ^{ \prime} u_{i}^{c}+
\bar{Q}_{3} \overline{ \eta}^{\prime } \bar{u}_{i}^{c} 
\right) + \frac{\kappa _{5\alpha i}}{3} \left( Q_{ \alpha} \rho u_{i}^{c}+
\bar{Q}_{ \alpha} \overline{ \rho} \bar{u}_{i}^{c} \right) \right],
\label{newint}
\end{eqnarray}
which imply a general mixing in the u-quark sector. Denoting
\begin{equation}
\begin{array}{c}
\psi _{u}^{+}= \left( \begin{array}{ccc} u_{1}&
                                         u_{2}&
                                         u_{3}
\end{array} \right)^{T}, \,\
\psi _{u}^{-}= \left( \begin{array}{ccc} u^{c}_{1}&
                                         u^{c}_{2}&
                                         u^{c}_{3}
\end{array} \right)^{T},
\end{array}
\label{cbasis}
\end{equation}
where all the u-quarks fields are still Weyl spinors, we can also, 
define \newline $\Psi _{u}^{\pm }=(\psi _{u}^{+}  \psi _{u}^{-}) ^{T}.$
We can define the mass term
$-(1/2)[ \Psi _{u}^{\pm T}Y_{u}^{ \pm}\Psi _{u}^{\pm }+H.c.]$
where $Y_{u}^{\pm}$ is given by: 
\begin{equation}
Y_{u}^{ \pm}=
\left( 
\begin{array}{cc}
0 & X_{u}^{T} \\ 
X_{u} & 0
\end{array}
\right),
\label{ypm}
\end{equation}
with
\begin{equation}
X_{u}=\frac{1}{3}\left( 
\begin{array}{ccc}
\kappa_{511}u & \kappa_{521}u & \kappa_{11}v^{\prime } \\
\kappa_{512}u & \kappa_{522}u & \kappa_{12}v^{\prime } \\
\kappa_{513}u & \kappa_{523}u & \kappa_{13}v^{\prime } \\
\end{array}
\right),
\label{cumm1}
\end{equation}
where the VEVs are defined in Eq.(\ref{vev1})

The u-quarks mass matrix $Y_{u}^{ \pm}$ is diagonalized using two rotation
matrices, $D$ and $E$, defined by
\begin{eqnarray}
u_{i}^{+}=D_{ij}\psi _{uj}^{+}, \,\
u_{i}^{-}=E_{ij}\psi _{uj}^{-}, \,\ i,j=1,2,3.
\label{2sc} 
\end{eqnarray}
Then we can write the diagonal matrix ($D$ and $E$ are unitary) as
\begin{equation}
M_{u}=E^{*}X_{u}D^{-1}.
\label{m1}
\end{equation}
To determine $D$ and $E$, we note that
\begin{equation}
M_{u}^{2}=DX_{u}^{T}X_{u}D^{-1}=E^{*}X_{u}X_{u}^{T}(E^{*})^{-1},
\label{m2}
\end{equation}
and define the following Dirac spinors
\begin{eqnarray}
\Psi (u^{+})=\left( \begin{array}{cc}
u^{+} & \bar{u}^{-} \end{array} \right)^{T}, \,\ 
\Psi ^{c}(u^{-})=\left( \begin{array}{cc}
u^{-} & \bar{u}^{+}
\end{array}
\right) ^{T}.
\label{emasssim}
\end{eqnarray}

To the d-quark type.There are interactions like
\begin{eqnarray}
- \left[ \frac{ \kappa_{2i}}{3} \left( Q_{3} \rho^{ \prime}
d_{i}^{c}+ \bar{Q}_{3} \overline{\rho }^{ \prime} \bar{d}_{i}^{c} \right) 
+ \frac{\kappa _{4\alpha i}}{3} \left( Q_{ \alpha} \eta d_{i}^{c}
+ \bar{Q}_{ \alpha} \overline{ \eta} \bar{d}_{i}^{c} \right) \right],
\end{eqnarray}
which imply a general mixing in the d-quark sector. Denoting
\begin{equation}
\begin{array}{c}
\psi _{d}^{+}= \left( \begin{array}{ccc} d^{c}_{1}&
                                         d^{c}_{2}&
                                         d^{c}_{3}
\end{array} \right)^{T}, \,\
\psi _{d}^{-}= \left( \begin{array}{ccc} d_{1}&
                                         d_{2}&
                                         d_{3}
\end{array} \right)^{T},
\end{array}
\label{cbasisd}
\end{equation}
where all the d-quarks fields are still Weyl spinors, we can also, 
define \newline $\Psi _{d}^{\pm }=(\psi _{d}^{+}  \psi _{d}^{-})^{T}.$
We can define the mass term
$-(1/2)[ \Psi _{d}^{\pm T}Y_{d}^{ \pm}\Psi _{d}^{\pm }+H.c.]$
where $Y_{d}^{ \pm}$ is given by: 
\begin{equation}
Y_{d}^{ \pm}=
\left( 
\begin{array}{cc}
0 & X_{d}^{T} \\ 
X_{d} & 0
\end{array}
\right),
\label{ypmd}
\end{equation}
with
\begin{equation}
X_{d}=\frac{1}{3}\left( 
\begin{array}{ccc}
\kappa_{411}v & \kappa_{412}v & \kappa_{413}v \\ 
\kappa_{421}v & \kappa_{422}v & \kappa_{423}v \\
\kappa_{21}u^{\prime} & \kappa_{22}u^{\prime} & \kappa_{23}u^{\prime}
\end{array}
\right),
\label{cdmm}
\end{equation}
where all the VEVs are defined in Eq.(\ref{vev1})

The d-quarks mass matrix $Y_{d}^{ \pm}$ is diagonalized using two rotation
matrices, $F$ and $G$, defined by
\begin{eqnarray}
d_{i}^{+}=F_{ij}\psi _{dj}^{+}, \,\ 
d_{i}^{-}=G_{ij}\psi _{uj}^{-}, \,\ i,j=1,2,3.
\label{2scd} 
\end{eqnarray}
Then we can write the diagonal matrix ($F$ and $G$ are unitary) as
\begin{equation}
M_{d}=G^{*}X_{d}F^{-1}.
\label{m1d}
\end{equation}
To determine $F$ and $G$, we note that
\begin{equation}
M_{d}^{2}=FX_{d}^{T}X_{d}F^{-1}=G^{*}X_{d}X_{d}^{T}(G^{*})^{-1},
\label{m2}
\end{equation}
and define the following Dirac spinors
\begin{eqnarray}
\Psi (d^{+})=\left( \begin{array}{cc}
d^{+} & \bar{d^{-}}
\end{array}
\right) ^{T}, \,\ 
\Psi ^{c}(d^{-})=\left( \begin{array}{cc}
d^{-} & \bar{d^{+}}
\end{array}
\right) ^{T}.
\label{emasssim}
\end{eqnarray}

In general the Yukawa couplings $\kappa _{1i}, \kappa _{5 \alpha i}, \kappa _{2i}$ and $\kappa _{4 \alpha i}$ are 
different of zero and all quarks get their masses as happen in the SM and in the MSSM. However, we know 
that the quarks $s,t$ are heavier then the quark $c,b$ and the quark $u$ is lighter than the quark $d$. 

We can try to given an reasonable explanation about these mass hierarchy 
in this model. In order to get this explanation we can suppose that the Yukawa couplings 
${\cal O}(\kappa _{11})={\cal O}(\kappa _{12})={\cal O}(\kappa _{21})={\cal O}(\kappa _{22})$ are much 
smaller than the other Yukawa couplings that appear at mass matrices of the usual quarks. This hypothesis  means we are 
going to neglect the mixing between the first and second familly with the first familly. 

Under this supposition we can rewrite our mass matrices, given at Eqs.(\ref{cumm1},\ref{cdmm}), in the following way
\begin{eqnarray}
X_{u}&\simeq&\frac{1}{3}\left( 
\begin{array}{cc}
\kappa_{511} & \kappa_{521}  \\
\kappa_{512} & \kappa_{522}  
\end{array}
\right) u,\nonumber \\
X_{d}&\simeq&\frac{1}{3}\left( 
\begin{array}{cc}
\kappa_{411} & \kappa_{412}  \\ 
\kappa_{421} & \kappa_{422} 
\end{array}
\right) v
\label{massaproxx}
\end{eqnarray}
We know from Eq.(\ref{valoresvev}) that $u > v$ then Eq.(\ref{massaproxx}) explain why the quarks of charge $(2e)/3$ 
are heavier than the quarks of charge $(-e)/3$. 

By another way 
\begin{eqnarray}
m_{u}= \frac{1}{3}\kappa _{13}v^{\prime}, \,\ m_{d}= \frac{1}{3}\kappa _{23}u^{\prime},
\end{eqnarray}
but, from Eq.(\ref{valoresvev}) we notice that $v^{\prime}\sim u^{\prime}$ and if $\kappa _{23}>\kappa _{13}$ we can 
explain why $d$-quark is a little more heavier than $u$-quark.

There are another way to try to explain the mass hierarchy between the fermions. If we look from 
Eqs.(\ref{clmm},\ref{cumm1},\ref{cdmm}), it is easy to note that we can prevent $u,d,s$ and $e$ from picking up 
tree-level masses. To get this result we need to impose the following ${\cal Z}^{\prime}_{2}$ symmetry on the Lagrangian
\begin{equation}
\widehat{d}_{2}^{c}\longrightarrow -\widehat{d}_{2}^{c},\,\ 
\widehat{d}_{3}^{c}\longrightarrow -\widehat{d}_{3}^{c},\,\ 
\widehat{u}_{3}^{c}\longrightarrow -\widehat{u}_{3}^{c},\,\
\widehat{l}_{3}^{c}\longrightarrow -\widehat{l}_{3}^{c},
\label{z2def}
\end{equation}
the others superfields are even under this symmetry as showed at Ref.~\cite{cmmc}. On this case, 
it was showed that under ${\cal Z}^{\prime}_{2}$ symmetry the heavy quarks $c,b$ and $t$ acquire 
mass at tree level while the light quarks ($u,d,s$) get their mass at 1-loop level. 

There is another intersting possibility to get $u$, $d$ light. We can introduce a new discrete $T^{\prime}$ flavor 
symmetry as done in \cite{Sen:2007vx}.
 
\subsection{Masses of Exotic quarks}

For the J-quark type. There are interactions like
\begin{equation}
- \left[ \frac{\kappa_{3}}{3} \left( Q_{3} \chi^{\prime }J^{c}+
\bar{Q}_{3} \overline{\chi^{\prime }} \bar{J}^{c} \right) \right],
\end{equation}
which imply one diagonalized state with the following mass
\begin{equation}
M_{J}^{mass}=-\frac{\kappa_{3}}{3} \omega^{\prime } \left( JJ^{c}+
\bar{J}\bar{J^{c}} \right).
\label{Jmass}
\end{equation}

The another exotic quark j. There are interactions like
\begin{equation}
- \left[ \frac{\kappa _{6\alpha \beta }}{3} \left( Q_{\alpha }\chi
j_{\beta }^{c}+\bar{Q}_{ \alpha} \overline{ \chi} \bar{j}_{ \beta}^{c} 
\right) \right],
\end{equation}
which imply a general mixing in the j-quark sector. Denoting
\begin{equation}
\begin{array}{c}
\psi _{j}^{+}=\left( \begin{array}{cc}
j_{1}^{c} & j_{2}^{c}
\end{array}
\right) ^{T}, \,\
\psi _{j}^{-}=\left( \begin{array}{cc}
j_{1} & j_{2}
\end{array}
\right) ^{T},
\end{array}
\label{cbasisj}
\end{equation}
where all the j-quarks fields are still Weyl spinors, we can also, 
define \newline $\Psi _{j}^{\pm }=\left(\psi _{j}^{+},  \psi _{j}^{-} \right) ^{T}.$
We can define the mass term
$-(1/2)\Psi _{j}^{\pm T}Y_{j}^{ \pm}\Psi _{j}^{\pm }+H.c.]$
where $Y_{j}^\pm$ is given by: 
\begin{equation}
Y_{j}^{ \pm}=
\left( 
\begin{array}{cc}
0 & X_{j}^{t} \\ 
X_{j} & 0
\end{array}
\right),
\label{ypm}
\end{equation}
with
\begin{equation}
X_{j}=\frac{1}{3}\left( 
\begin{array}{cc}
\kappa _{611}w & \kappa _{612}w \\ 
\kappa _{621}w & \kappa _{622}w
\end{array}
\right),
\label{clmm1}
\end{equation}
where the values VEVs are defined in Eq.(\ref{vev1}).

The j-quarks mass matrix is diagonalized using two rotation matrices, 
$H$ and $I$, defined by
\begin{eqnarray}
j_{\alpha }^{+}=H_{\alpha \beta }\psi _{j_{\beta }}^{+}, \,\
j_{\alpha }^{-}=I_{\alpha \beta }\psi _{j_{\beta }}^{-}, \,\  \alpha ,\beta
=1,2.
\label{2scj} 
\end{eqnarray}
Then we can write the diagonal matrix ($H$ and $I$ are unitary) as
\begin{equation}
M_{j}=I^{*}X_{j}H^{-1}.
\label{m1j}
\end{equation}
To determine $I$ and $H$, we note that
\begin{equation}
M_{j}^{2}=HX_{j}^{T}X_{j}H^{-1}=I^{*}X_{j}X_{j}^{T}(I^{*})^{-1}.
\label{m2j}
\end{equation}

The masses of physical $j$ are
\begin{eqnarray}
M_{j_{1}}&=& \frac{1}{6} \left( \kappa _{611}+ \kappa _{622}+\sqrt{(\kappa _{611}- \kappa _{622})^{2}+
4\kappa _{612}\kappa _{621}} \right)w , \nonumber \\
M_{j_{2}}&=& \frac{1}{6} \left( \kappa _{611}+ \kappa _{622}-\sqrt{(\kappa _{611}- \kappa _{622})^{2}+
4\kappa _{612}\kappa _{621}} \right)w .
\end{eqnarray}

Notice that if $\kappa _{612}$ is zero we get that 
\begin{equation}
M_{j_{1}}=\frac{\kappa _{611}}{3}w, \,\ \mbox{ and } \,\ 
M_{j_{2}}=\frac{\kappa _{622}}{3}w,
\label{jmasses}
\end{equation} 
therefore $\kappa _{612}$ and $\kappa _{621}$ can be both zero that both $j_{1}$ and $j_{2}$ are massive. 

If we consider that ${\cal O}(\kappa_{3})$, ${\cal O}(\kappa_{611})$ and ${\cal O}(\kappa_{622})$ are the same order, 
for example ${\cal O}\simeq 10^{-2}$ for example, it means from Eqs.(\ref{valoresvev},\ref{Jmass},\ref{jmasses}) that 
$J$-quark is heavier than $j_{1,2}$ quarks and their masses are in TeV scale.

We define the following Dirac spinors
\begin{eqnarray}
\Psi (j^{+})=\left( \begin{array}{cc}
j^{+} & \bar{j^{-}}
\end{array}
\right) ^{T}, \,\ 
\Psi ^{c}(j^{-})=\left( \begin{array}{cc}
j^{-} & \bar{j^{+}}
\end{array}
\right) ^{T}.
\label{emasssimj}
\end{eqnarray}

\subsection{The masses of Gluinos}

It is well known gluinos are the supersymmetric partners of the gluons.
Therefore gluinos, as in the MSSM, are the color octet fermions in the model and due the fact that the $SU(3)_{c}$ group 
is unbroken, it means the gluinos can not mix with any others particles in the model, then they are already 
mass eigenstates.

Their mass, is one of the soft parameter that break SUSY, can be written as 
\begin{equation}
\mathcal{L}_{\mathrm{mass}}^{\mathrm{gluino}}=\frac{m_{\tilde{g}}}{2}
\bar{\tilde{g}}\tilde{g}
\end{equation}
so that its mass at tree level is $m_{\tilde{g}}=|m_{\lambda_c}|$, as denoted at Eq.(\ref{gmt}) and 
Refs.~\cite{331susy1,mcr}, where 
\begin{equation}
\tilde{g}^{a}=\left( 
\begin{array}{c}
- \imath \lambda _{C}^{a} \\ 
\imath \overline{\lambda _{C}^{a}}
\end{array}
\right) \,\ ,\hspace{1cm}a=1,\ldots ,8, 
\end{equation}
is the Majorana four-spinor defining the physical gluinos states.

\subsection{Gauge Boson´s masses}

The gauge mass term is given by ${\cal L}^{\mbox{scalar}}_{HHVV}$ which we 
can divided in ${\cal L}_{\mbox{mass}}^{\mbox{charged}}$ and 
${\cal L}_{\mbox{mass}}^{\mbox{neutral}}$, see Refs.~\cite{331susy1,mcr}.

The neutral gauge boson mass is given by
\begin{eqnarray}
{\cal L}_{\mbox{mass}}^{\mbox{neutral}}&=& \left( \begin{array}{ccc}
V_{3m}& V_{8m}& V_{m} \end{array} \right) M^{2} \left( \begin{array}{c}
V_{3}^{m} \\
V_{8}^{m} \\
V^{m} 
\end{array} \right),
\label{numerar}
\end{eqnarray}
where
\begin{eqnarray}
M^{2}= \frac{g^{2}}{2} \left( \begin{array}{ccc}
(v^{2}+u^{2})& \frac{1}{ \sqrt{3}}(v^{2}-u^{2})& 
-2tu^{2} \\
\frac{1}{ \sqrt{3}}(v^{2}-u^{2})& \frac{1}{3}(v^{2}+u^{2}+4w^{2})& 
\frac{2t}{ \sqrt{3}}(u^{2}+2w^{2}) \\
-2tu^{2}& \frac{2t}{ \sqrt{3}}(u^{2}+2w^{2})& 4t^{2} (u^{2}+w^{2})
\end{array} \right), 
\label{numerar}
\end{eqnarray}
with $t=g^{\prime}/g$.

In the approximation that $w^{2} \gg v^{2},u^{2}$, the masses of the neutral 
gauge bosons are: $0$, 
$M_{Z}^{2}$ and $M_{Z^{ \prime}}^{2}$, and the masses are given by
\begin{eqnarray}
M_{Z}^{2} \approx \frac{1}{4} \frac{g^{2}+4g^{ \prime 2}}{g^{2}+3g^{ \prime 2}}
(v_{\eta}^{2}+v_{\rho}^{2}+v_{\eta^{\prime}}^{2}+v_{\rho^{\prime}}^{2}), \,\
M_{Z^{ \prime}}^{2} \approx \frac{1}{3}(g^{2}+3g^{\prime 2})(v_{\chi}^{2}+v_{\chi^{\prime}}^{2}).
\label{zmass}
\end{eqnarray}

The charged gauge boson mass term, ${\cal L}_{\mbox{mass}}^{\mbox{neutral}}$ see Refs.~\cite{331susy1,mcr}, can be 
written as
\begin{eqnarray}
{\cal L}_{\mbox{mass}}^{\mbox{charged}}
&=& M^{2}_{W} W^{-}_{m}W^{+ m}+ M^{2}_{V} V^{-}_{m} V^{+ m}+ 
M^{2}_{U} U^{--}_{m}U^{++ m},
\label{bgmcc}
\end{eqnarray}
where 
\begin{eqnarray}
M^{2}_{U}&=& \frac{g^{2}}{4}(v^{2}_{ \rho}+v^{2}_{ \chi}+
v^{2}_{ \rho^{\prime}}+v^{2}_{ \chi^{\prime}}), 
\nonumber \\ 
M^{2}_{W}&=& \frac{g^{2}}{4}(v^{2}_{ \eta}+v^{2}_{ \rho}+
v^{2}_{ \eta^{\prime}}+v^{2}_{ \rho^{\prime}}), \nonumber \\
M^2_{V}&=& \frac{g^{2}}{4}(v^{2}_{ \eta}+v^{2}_{ \chi}+
v^{2}_{ \eta^{\prime}}+v^{2}_{ \chi^{\prime}}).
\label{wmass}
\end{eqnarray}
Comparing Eqs.(\ref{wmass},\ref{zmass}) we can conclude
\begin{equation}
M_{Z^{\prime}}>M_{U}>M_{V}>M_{Z}>M_{W}.
\label{massasmaiores}
\end{equation}

Using $M_W$ given in Eq.(\ref{wmass}) and $M_Z$ in Eq.(\ref{zmass}) we get
 the following relation:
\begin{eqnarray}
\frac{M^2_Z}{M^2_W}= \frac{1+4t^2}{1+3t^2}=\frac{1}{1- \sin^2 \theta_W},
\label{rel1}
\end{eqnarray}
therefore we obtain
\begin{equation}
t^2= \frac{ \sin^2 \theta_W}{1-4 \sin^2 \theta_W}.
\label{tdef}
\end{equation}
We want to mention that the gauge boson sector is exactly the same as in the non-supersymmetric 
3-3-1 model.

Using Eq.(\ref{valoresvev}) in Eqs.(\ref{zmass},\ref{wmass}) we get the following masses values for the gauge bosons
\begin{eqnarray}
M_{U}&=&734,63 \,\ \mbox{GeV}, \,\ M_{V}=730,28 \,\ \mbox{GeV}, \,\ M_{W}=80,40 \mbox{GeV}, \nonumber \\ 
M_{Z}&=&91,3 \,\ \mbox{GeV}, \,\ M_{Z^{\prime}}=2698,94 \,\ \mbox{GeV}. 
\end{eqnarray}
These values satisfy the Eq.(\ref{massasmaiores}) and $M_{W}$ and $M_{Z}$ are in agreement with the experimental limits. 
The lower limit in $Z^{\prime}$ boson is $M_{Z^{\prime}}>822$ GeV \cite{pdg} and our mass is in agreement with this 
experimental limit.

\section{Conclusions}
\label{sec:conclusion}

In this paper we have presented new  $R$-symmetry  for the
minimal supersymmetric $\mbox{SU}(3)_{C}\otimes \mbox{SU}(3)_{L} \otimes \mbox{U}(1)_{X}$  model and studied
all the spectrum from the fermion's sector and gauged's boson sector of this model. We also show that some of 
the new state as $\tilde{\chi}_{1 \ldots 4}^{\pm \pm}$, $\tilde{\chi}_{1 \ldots 7}^{\pm \pm}$, 
$\tilde{\chi}_{1 \pm 9}^{0}$, $j_{1}$ and $J$ can be discovered by LHC or same ILC, if they really exist.

The new $R$-parity not only provides a simple mechanism
for the mass generation of the neutrinos but also gives  some
lepton flavor violating  interactions at the tree level. This will
play some important phenomenology in our model such as the
proton's stability, forbiddance of the neutron-antineutron
oscillation and neutrinoless double beta decay. 

\begin{center}
{\bf Acknowledgments} 
\end{center}
This work was financed by the Brazilian funding agency
CNPq, under contract number 309564/2006-9. 

\appendix

\section{Lagrangian}
\label{sec:lagrangian}

The goal of this Appendix is to present all terms in the lagrangian of the model, which we 
have used in this work.

\subsection{Lepton Lagrangian}
\label{a1}

In the ${\cal L}_{\mbox{Lepton}}$ term can be written as \cite{wb}
\begin{eqnarray}
{\cal L}_{\mbox{L\'epton}}
&=& {\cal L}_{llV}+ {\cal L}_{ \tilde{l} \tilde{l}V}+ 
{\cal L}_{l \tilde{l} \tilde{V}}+ {\cal L}_{ \tilde{l} \tilde{l}VV}
+{\cal L}^{leptons}_{cin}+ 
{\cal L}^{leptons}_{F}+ {\cal L}^{leptons}_{D}, \nonumber \\
\label{lint}
\end{eqnarray}
where
\begin{eqnarray}
{\cal L}_{llV}&=&\frac{g}{2} \bar{L}\bar\sigma^m\lambda^a LV^a_m; \,\
{\cal L}_{ \tilde{l} \tilde{l}V}=- \frac{ig}{2}\left[ 
\tilde{L}\lambda^a\partial^m\bar{\tilde{L}}- 
\bar{\tilde{L}}\lambda^a\partial^m \tilde{L} \right]V^{a}_{m}, \nonumber \\
{\cal L}_{l \tilde{l} \tilde{V}}&=&- \frac{ig}{ \sqrt{2}}
( \bar{L}\lambda^a\tilde{L}\bar{\lambda}^a_{A}- 
\bar{\tilde{L}}\lambda^aL\lambda^a_{A}); \,\
{\cal L}_{ \tilde{l} \tilde{l}VV}= \frac{g^2}{4} V_m^aV^{bm}
\bar{\tilde{L}} \lambda^{a}\lambda^{b} \tilde{L}, \nonumber \\
{\cal L}^{leptons}_{cin}&=& \tilde{L} \partial^{2} \tilde{L}^{*}- 
iL \sigma^m \partial_m \bar{L}; \,\  \partial^{2}= \partial^{m} \partial_{m}, \nonumber \\
{\cal L}^{leptons}_{F}&=& \vert F_{L} \vert^2; \nonumber \\
{\cal L}^{leptons}_{D}&=& \frac{g}{2} \bar{\tilde{L}}\lambda^a\tilde{L} D^a. \nonumber \\
\label{lepbos}
\end{eqnarray}

\subsection{Quark Lagrangian}
\label{a2}

As above we can write to the quarks
\begin{eqnarray}
{\cal L}_{qqV}&=& \frac{g_{s}}{2} ( 
\bar{Q}_{i}\bar\sigma^m\lambda^a Q_{i}- 
\bar{u}^{c}_i\bar\sigma^m\lambda^{*a} u^{c}_i-
\bar{d}^{c}_i\bar\sigma^m\lambda^{*a} d^{c}_i-
\bar{J}^{c}_i\bar\sigma^m\lambda^{*a} J^{c}_i)g^a_m  \nonumber \\ 
&+& \frac{g}{2}( \bar{Q}_{3}\bar\sigma^m\lambda^a Q_{3}- 
\bar{Q}_{ \alpha}\bar\sigma^m\lambda^{*a} Q_{ \alpha})V^a_m  \nonumber \\
&+& \frac{g^{ \prime}}{2} \left( 
\frac{2}{3} \bar{Q}_{3}\bar\sigma^m Q_{3}- 
\frac{1}{3} \bar{Q}_{ \alpha}\bar\sigma^m Q_{ \alpha}-
\frac{2}{3} \bar{u}^{c}_i\bar\sigma^m u^{c}_i+
\frac{1}{3} \bar{d}^{c}_i\bar\sigma^m d^{c}_i-
\frac{5}{3} \bar{J}^{c}\bar\sigma^m J^{c}+
\frac{4}{3} \bar{j}^{c}_{ \beta}\bar\sigma^m j^{c}_{ \beta} 
\right)V_m, \nonumber \\
{\cal L}_{ \tilde{q} \tilde{q}V}&=& \frac{-ig_{s}}{2} \left[ (
\tilde{Q}_{i}\lambda^a\partial^m\bar{\tilde{Q}}_{i}-
\bar{\tilde{Q}}_{i}\lambda^a\partial^m \tilde{Q}_{i}-
\tilde{u}^{c}_i\lambda^{*a}\partial^m\bar{\tilde{u}}^{c}_i + 
\bar{\tilde{u}}^{c}_i\lambda^{*a}\partial^m \tilde{u}^{c}_i \right. \nonumber \\
&-& \left.
\tilde{d}^{c}_i\lambda^{*a}\partial^m\bar{\tilde{d}}^{c}_i+
\bar{\tilde{d}}^{c}_i\lambda^{*a}\partial^m \tilde{d}^{c}_i -
\tilde{J}^{c}_i\lambda^{*a}\partial^m\bar{\tilde{J}}^{c}_i+
\bar{\tilde{J}}^{c}_i\lambda^{*a}\partial^m \tilde{J}^{c}_i)g^a_m \right] \nonumber \\ 
&-& \frac{ig}{2} ( \tilde{Q}_{3}\lambda^a\partial^m\bar{\tilde{Q}}_{3}-
\bar{\tilde{Q}}_{3}\lambda^a\partial^m \tilde{Q}_{3}-
\tilde{Q}_{ \alpha}\lambda^{*a}\partial^m\bar{\tilde{Q}}_{ \alpha}+
\bar{\tilde{Q}}_{ \alpha}\lambda^{*a}\partial^m \tilde{Q}_{ \alpha})V^a_m \nonumber \\
&-& \frac{ig^{ \prime}}{2} \left[ \frac{2}{3} (
\tilde{Q}_{3} \partial^m\bar{\tilde{Q}}_{3}-
\bar{\tilde{Q}}_{3} \partial^m \tilde{Q}_{3})- \frac{1}{3} (
\tilde{Q}_{ \alpha} \partial^m\bar{\tilde{Q}}_{ \alpha}-
\bar{\tilde{Q}}_{ \alpha} \partial^m \tilde{Q}_{ \alpha})-
\frac{2}{3} ( \tilde{u}^{c}_i \partial^m\bar{\tilde{u}}^{c}_i-
\bar{\tilde{u}}^{c}_i \partial^m \tilde{u}^{c}_i) \right. \nonumber \\
&+& \left. \frac{1}{3}
(\tilde{d}^{c}_i \partial^m\bar{\tilde{d}}^{c}_i-
\bar{\tilde{d}}^{c}_i \partial^m \tilde{d}^{c}_i)- \frac{5}{3}
(\tilde{J}^{c} \partial^m\bar{\tilde{J}}^{c}-
\bar{\tilde{J}}^{c} \partial^m \tilde{J}^{c})+ \frac{4}{3}
(\tilde{j}^{c}_{ \beta} \partial^m\bar{\tilde{j}}^{c}_{ \beta}-
\bar{\tilde{j}}^{c}_{ \beta} \partial^m \tilde{j}^{c}_{ \beta}) \right] V_m, \nonumber \\
{\cal L}_{q \tilde{q} \tilde{V}}&=& \frac{-ig_{s}}{ \sqrt{2}} \left[ ( 
\bar{Q}_{i}\lambda^a\tilde{Q}_{i}-
\bar{u}^{c}_i\lambda^{*a}\tilde{u}^{c}_i- \bar{d}^{c}_i\lambda^{*a}\tilde{d}^{c}_i-
\bar{J}^{c}_i\lambda^{*a}\tilde{J}^{c}_i) \bar{\lambda}^a_{c} \right. \nonumber \\
&-& \left. ( \bar{\tilde{Q}}_{i}\lambda^aQ_{i}-
\bar{\tilde{u}}^{c}_i\lambda^{*a}u^{c}_i-
\bar{\tilde{d}}^{c}_i\lambda^{*a}d^{c}_i- \bar{\tilde{J}}^{c}_i\lambda^{*a}J^{c}_i)
\lambda^a_{c} \right] \nonumber \\
&-& \frac{ig}{ \sqrt{2}} \left[ ( 
\bar{Q}_{3}\lambda^a\tilde{Q}_{3}- \bar{Q}_{\alpha}\lambda^{*a}\tilde{Q}_{\alpha})\bar{\lambda}^a_{A}- ( \bar{\tilde{Q}}_{3}\lambda^aQ_{3}-
\bar{\tilde{Q}}_{\alpha}\lambda^{*a}Q_{\alpha}) \lambda^a_{A} \right] \nonumber \\
&-& \frac{ig'}{\sqrt{2}} \left[ \left( \frac{2}{3} \bar{Q}_{3}\tilde{Q}_{3}- 
\frac{1}{3}\bar{Q}_{\alpha}\tilde{Q}_{\alpha}-\frac{2}{3} \bar{u}^{c}_i\tilde{u}^{c}_i+
\frac{1}{3} \bar{d}^{c}_i\tilde{d}^{c}_i-\frac{5}{3} \bar{J}^{c}\tilde{J}^{c}+
\frac{4}{3} \bar{j}^{c}_{ \beta}\tilde{j}^{c}_{ \beta} \right) \bar{\lambda}_{B} 
\right. \nonumber \\
&-& \left. \left( 
\frac{2}{3}\bar{\tilde{Q}}_{3}Q_{3}- \frac{1}{3}\bar{\tilde{Q}}_{\alpha}Q_{\alpha}-
\frac{2}{3} \bar{\tilde{u}}^{c}_iu^{c}_i+\frac{1}{3} \bar{\tilde{d}}^{c}_id^{c}_i-
\frac{5}{3} \bar{\tilde{J}}^{c}J^{c}+
\frac{4}{3} \bar{\tilde{j}}^{c}_{ \beta}j^{c}_{ \beta} \right) \lambda_B \right], 
\nonumber \\
{\cal L}_{ \tilde{q} \tilde{q}VV}&=& \frac{-1}{4} \left[ g_{s}^2(
\bar{\tilde{Q}}_{i}\lambda^{a}\lambda^{b} \tilde{Q}_{i}+
\bar{\tilde{u}}^{c}_i\lambda^{*a}\lambda^{*b} \tilde{u}^{c}_i+
\bar{\tilde{d}}^{c}_i\lambda^{*a}\lambda^{*b} \tilde{d}^{c}_i+
\bar{\tilde{J}}^{c}_i\lambda^{*a}\lambda^{*b} \tilde{J}^{c}_i)g^a_mg^{bm} 
\right] \nonumber \\
&-& \frac{1}{4} \left[ g^2(
\bar{\tilde{Q}}_{3}\lambda^{a}\lambda^{b} \tilde{Q}_{3}+
\bar{\tilde{Q}}_{ \alpha}\lambda^{*a}\lambda^{*b} \tilde{Q}_{ \alpha}) 
\right] V^a_mV^{bm} 
- \frac{1}{2} \left[ g_{s}g(
\bar{\tilde{Q}}_{3}\lambda^{a}\lambda^{b} \tilde{Q}_{3}+
\bar{\tilde{Q}}_{ \alpha}\lambda^{a}\lambda^{*b} \tilde{Q}_{ \alpha}) 
\right] g^a_mV^{bm} \nonumber \\
&-&  \frac{1}{2}g_{s}g^{ \prime} \left[ 
\frac{2}{3} \bar{\tilde{Q}}_{3}\lambda^a \tilde{Q}_{3}-
\frac{1}{3} \bar{\tilde{Q}}_{ \alpha}\lambda^a \tilde{Q}_{ \alpha}+
\frac{2}{3} \bar{\tilde{u}}^{c}_i\lambda^{*a} \tilde{u}^{c}_i-
\frac{1}{3} \bar{\tilde{d}}^{c}_i\lambda^{*a} \tilde{d}^{c}_i+
\frac{5}{3} \bar{\tilde{J}}^{c}\lambda^{*a} \tilde{J}^{c} \right. \nonumber \\
&-& \left.
\frac{4}{3} \bar{\tilde{j}}^{c}_{ \beta}\lambda^{*a} \tilde{j}^{c}_{ \beta} \right] 
g^{am}V_m-  \frac{1}{2}gg^{ \prime} \left[ 
\frac{2}{3} \bar{\tilde{Q}}_{3}\lambda^a \tilde{Q}_{3}+
\frac{1}{3} \bar{\tilde{Q}}_{ \alpha}\lambda^{*a} \tilde{Q}_{ \alpha} \right] 
V^{am}V_m \nonumber \\
&-& \frac{1}{4} g^{ \prime 2} \left[ 
\frac{4}{9}(\bar{\tilde{Q}}_{3}\tilde{Q}_{3}+\bar{\tilde{u}}^{c}_i\tilde{u}^{c}_i)+
\frac{1}{9}(\bar{\tilde{Q}}_{\alpha}\tilde{Q}_{\alpha}+
\bar{\tilde{d}}^{c}_i\tilde{d}^{c}_i)+ \frac{25}{9} \bar{\tilde{J}}^{c}\tilde{J}^{c}+
\frac{16}{9} \bar{\tilde{j}}^{c}_{\beta}\tilde{j}^{c}_{\beta} \right] V^m V_m. \nonumber \\
{\cal L}^{quark}_{cin}&=&\tilde{Q}_i \partial^{2} \tilde{Q}^{*}_i+
\tilde{u}^{c}_i \partial^{2} \tilde{u}^{c*}_i+
\tilde{d}^{c}_i \partial^{2} \tilde{d}^{c*}_i+
\tilde{J}^{c}_i \partial^{2} \tilde{J}^{c*}_i- 
iQ_i \sigma^m \partial_m \bar{Q_i}-
iu^{c}_{i} \sigma^m \partial_m \bar{u}^{c}_{i} \nonumber \\
&-&id^{c}_{i} \sigma^m \partial_m \bar{d}^{c}_{i}-
iJ^{c}_{i} \sigma^m \partial_m \bar{J}^{c}_{i}, \nonumber \\
{\cal L}^{quark}_{F}&=& \vert F_{Q_i} \vert^2 +
 \vert F_{u_i} \vert^2 + \vert F_{d_i} \vert^2 + 
\vert F_{J_i} \vert^2, \nonumber \\
{\cal L}^{quark}_{D}&=& \frac{g_{s}}{2}( \bar{\tilde{Q}}_{i}\lambda^a\tilde{Q}_{i} -
\bar{\tilde{u}}^{c}_{i}\lambda^{*a}\tilde{u}^{c}_{i}- 
\bar{\tilde{d}}^{c}_{i}\lambda^{*a}\tilde{d}^{c}_{i}-
\bar{\tilde{J}}^{c}_{i}\lambda^{*a}\tilde{J}^{c}_{i})D^a_c 
+\frac{g}{2} \left( \bar{\tilde{Q}}_{3}\lambda^a\tilde{Q}_{3}-
\bar{\tilde{Q}}_{ \alpha}\lambda^{*a}\tilde{Q}_{ \alpha}  \right) D^{a} \nonumber \\
&+& \frac{g^{\prime}}{2} \left[ \frac{2}{3} \bar{\tilde{Q}}_{3}\tilde{Q}_{3} -
 \frac{1}{3} \bar{\tilde{Q}}_{ \alpha}\tilde{Q}_{ \alpha}
- \frac{2}{3} \bar{\tilde{u}}^{c}_{i}\tilde{u}^{c}_{i}+ 
\frac{1}{3} \bar{\tilde{d}}^{c}_{i}\tilde{d}^{c}_{i}-
\frac{5}{3} \bar{\tilde{J}}^{c}\tilde{J}^{c}+
\frac{4}{3} \bar{\tilde{j}}^{c}_{ \beta}\tilde{j}^{c}_{ \beta}
 \right] D. \nonumber \\
\label{quarkbos}
\end{eqnarray}

\subsection{Scalar Lagrangian}
\label{a3}

\begin{eqnarray}
{\cal L}^{Escalar}_{F}&=& \vert F_{\eta} \vert^2+ \vert F_{\rho} \vert^2+ 
\vert F_{\chi} \vert^2+  
\vert F_{\eta^{\prime}} \vert^2+ 
\vert F_{\rho^{\prime}} \vert^2+ 
\vert F_{\chi^{\prime}} \vert^2, \nonumber \\
{\cal L}^{Escalar}_{D}&=& \frac{g}{2} \left[ \bar{\eta}\lambda^a\eta+ 
\bar{\rho}\lambda^a\rho+ \bar{\chi}\lambda^a\chi+ 
\bar{\eta}^{\prime}\lambda^{* a}\eta^{\prime}- 
\bar{\rho}^{\prime}\lambda^{* a}\rho^{\prime}- 
\bar{\chi}^{\prime}\lambda^{* a}\chi^{\prime} \right] D^{a} \nonumber \\
&+& \frac{g^{ \prime}}{2} \left[  
\bar{\rho}\rho- \bar{\chi}\chi- \bar{\rho}^{\prime}\rho^{\prime}+ 
\bar{\chi}^{\prime}\chi^{\prime} \right]D, \nonumber \\
{\cal L}_{Higgs}&=& ({\cal D}_{m} \eta)^{\dagger}({\cal D}^{m} \eta)+ 
({\cal D}_{m} \rho)^{\dagger}({\cal D}^{m} \rho)+
({\cal D}_{m} \chi)^{\dagger}({\cal D}^{m} \chi)+
(\overline{{\cal D}_{m}} \eta^{\prime})^{\dagger}(\overline{{\cal D}^{m}} \eta^{\prime}) \nonumber \\ &+&
(\overline{{\cal D}_{m}} \rho^{\prime})^{\dagger}(\overline{{\cal D}^{m}} \rho^{\prime})+
(\overline{{\cal D}_{m}} \chi^{\prime})^{\dagger}(\overline{{\cal D}^{m}} \chi^{\prime})], \nonumber \\
{\cal L}_{Higgsinos}&=&
i \bar{\tilde{\eta}} \bar{\sigma}^{m}{\cal D}_{m}\tilde{\eta}+
i \bar{\tilde{\rho}} \bar{\sigma}^{m}{\cal D}_{m}\tilde{\rho}+
i \bar{\tilde{\chi}} \bar{\sigma}^{m}{\cal D}_{m}\tilde{\chi}+
i \bar{\tilde{\eta}}^{\prime} \bar{\sigma}^{m}\overline{{\cal D}_{m}}\tilde{\eta}^{\prime}+
i \bar{\tilde{\rho}}^{\prime} \bar{\sigma}^{m}\overline{{\cal D}_{m}}\tilde{\rho}^{\prime}+
i \bar{\tilde{\chi}}^{\prime} \bar{\sigma}^{m}\overline{{\cal D}_{m}}\tilde{\chi}^{\prime} ], \nonumber \\
{\cal L}_{H \tilde{H} \tilde{V}}&=&- \frac{ig}{ \sqrt{2}} \left[  
\bar{\tilde{\eta}}\lambda^a\eta\bar{\lambda}^a_{A} 
- \bar{\eta}\lambda^a\tilde{\eta}\lambda^a_{A}+
\bar{\tilde{\rho}}\lambda^a\rho\bar{\lambda}^a_{A} 
- \bar{\rho}\lambda^a\tilde{\rho}\lambda^a_{A}+
\bar{\tilde{\chi}}\lambda^a\chi\bar{\lambda}^a_{A} 
- \bar{\chi}\lambda^a\tilde{\chi}\lambda^a_{A}+
\bar{\tilde{\eta}}^{\prime}\lambda^{* a}\eta^{\prime}\bar{\lambda}^a_{A} 
\right. \nonumber \\
&+& \left. \bar{\eta}^{\prime}\lambda^{* a}\tilde{\eta}^{\prime}\lambda^a_{A}-
\bar{\tilde{\rho}}^{\prime}\lambda^{* a}\rho^{\prime}\bar{\lambda}^a_{A} 
+ \bar{\rho}^{\prime}\lambda^{* a}\tilde{\rho}^{\prime}\lambda^a_{A}-
\bar{\tilde{\chi}}^{\prime}\lambda^{* a}\chi^{\prime}\bar{\lambda}^a_{A} +
\bar{\chi}^{\prime}\lambda^{* a}\tilde{\chi}^{\prime}\lambda^a_{A} \right] \nonumber \\ &-&
\frac{ig^{ \prime}}{ \sqrt{2}} \left[  
\bar{\tilde{\rho}}\rho\bar{\lambda}_{B} 
-\bar{\rho}\tilde{\rho}\lambda_{B}-
\bar{\tilde{\chi}}\chi\bar{\lambda}_{B} + \bar{\chi}\tilde{\chi}\lambda_{B} 
- \bar{\tilde{\rho}}^{\prime}\rho^{\prime}\bar{\lambda}_{B} 
+\bar{\rho}^{\prime}\tilde{\rho}^{\prime}\lambda_{B}+
\bar{\tilde{\chi}}^{\prime}\chi^{\prime}\bar{\lambda}_{B} 
-\bar{\chi}^{\prime}\tilde{\chi}^{\prime}\lambda_{B}
\right], \nonumber \\
\label{mix1}
\end{eqnarray}
where the covariant derivative of $SU(3)$ are given by:
\begin{eqnarray}
{\cal D}_{m}\phi_{i}  &=&  \partial_{m}\phi_{i} - ig\left( \vec{V}_{m} 
.\frac{\vec{\lambda}}{2}\right)^{j}_{i}\phi_{j} - 
ig^\prime N_{\phi_{i}}V_{m}\phi_{i}, \nonumber \\
\overline{{\cal D}_{m}}\phi_{i}  &=&  \partial_{m}\phi_{i} - ig\left( \vec{V}_{m} 
.\frac{\vec{\overline{\lambda}}}{2}\right)^{j}_{i}\phi_{j} - 
ig^\prime N_{\phi_{i}}V_{m}\phi_{i},
\label{dcat}
\end{eqnarray}

\subsection{Gauge Lagrangian}
\label{a4}

Now we are dealing with ${\cal L}_{Gauge}$ that can be expanded as
\begin{eqnarray}
{\cal L}_{Gauge}&=& {\cal L}_{dc}+{\cal L}^{gauge}_{D}.
\end{eqnarray}
where
\begin{eqnarray}
{\cal L}^{gauge}_{D}&=&\frac{1}{2}D^{a}_{C}D^{a}_{C}+ \frac{1}{2}D^{a}D^{a}+ 
\frac{1}{2}DD, \nonumber \\
{\cal L}_{dc}&=&- \frac{1}{4} \left( G^{amn}G^{a}_{mn}+W^{amn}W^{a}_{mn}+F^{mn}F_{mn} \right) \nonumber \\
&-& \imath \left( \bar{\lambda}^{a}_{C} \bar{\sigma}^{n}{\cal D}^{C}_{n}\lambda^{a}_{C}+
\bar{\lambda}^{a}_{A} \bar{\sigma}^{n}{\cal D}^{L}_{n}\lambda^{a}_{A}+
\bar{\lambda}_{B}\bar{\sigma}^{n}\partial_{n}\lambda_{B} \right) \,\ , 
\label{tutty}
\end{eqnarray}
with
\begin{eqnarray}
G^{a}_{mn}&=& \partial_{m}g^{a}_{n}-\partial_{n}g^{a}_{m}-gf^{abc}g^{b}_{m}
g^{c}_{n}, \nonumber \\
W^{a}_{mn}&=& \partial_{m}V^{a}_{n}-\partial_{n}V^{a}_{m}-gf^{abc}V^{b}_{m}
V^{c}_{n}, \nonumber \\
B_{mn}&=& \partial_{m}V_{n}-\partial_{n}V_{m}, \nonumber \\
{\cal D}^{C}_{n}\lambda^{a}_{C}&=&\partial_{n}\lambda^{a}_{C}-g_{s}f^{abc}g^{b}_{m}\lambda^{c}_{C}, \nonumber \\
{\cal D}^{L}_{n}\lambda^{a}_{A}&=&\partial_{n}\lambda^{a}_{A}-gf^{abc}V^{b}_{m}\lambda^{c}_{A}, 
\label{gauge2}
\end{eqnarray}
$f^{abc}$ are the constant structure of $SU(3)$ gauge group.

\subsection{Superpotential Lagrangian}
\label{a5}

\begin{eqnarray}
W_{2}&=&{\cal L}^{W2}_{F}+{\cal L}_{ \tilde{ \eta}L}+
{\cal L}_{HMT}, \nonumber \\
W_{3}&=&{\cal L}^{W3}_{F}+{\cal L}_{ll \tilde{l}}+{\cal L}_{llH}+
{\cal L}_{l \tilde{l} \tilde{H}}+{\cal L}_{l\tilde{H}H}+
{\cal L}_{\tilde{l}HH}+{\cal L}_{H \tilde{H} \tilde{H}}+
{\cal L}_{qqH}+{\cal L}_{q \tilde{q} \tilde{H}} \nonumber \\
&+& {\cal L}_{lq \tilde{q}}+{\cal L}_{\tilde{l}q \tilde{q}}+
{\cal L}_{qq \tilde{q}},  
\end{eqnarray}
As componentes de cada lagrangiana s\~ao escritas como
\begin{eqnarray}
{\cal L}^{W2}_{F}&=& \frac{\mu_0}{2} \left( \tilde{L} F_{ \eta^{\prime}}+ 
\eta^{\prime} F_{L} \right)+
\frac{\mu_{ \eta}}{2} \left( \eta F_{\eta^{\prime}}+ \eta^{\prime} F_{ \eta} \right)+ 
\frac{\mu_{ \rho}}{2} \left( \rho F_{\rho^{\prime}}+ \rho^{\prime} F_{ \rho} \right)+ 
\frac{\mu_{ \chi}}{2} \left( \chi F_{\chi^{\prime}}+ \chi^{\prime} F_{ \chi} \right),
\nonumber \\
{\cal L}_{ \tilde{ \eta}L}&=&- \frac{\mu_0}{2}L \tilde{ \eta}^{\prime}; 
\,\
{\cal L}_{HMT}=- 
\frac{\mu_{ \eta}}{2} \tilde{ \eta}_i \tilde{ \eta}^{\prime}_i
-\frac{\mu_{ \rho}}{2} \tilde{ \rho}_i \tilde{ \rho}^{\prime}_i-
\frac{\mu_{ \chi}}{2} \tilde{ \chi}_i \tilde{ \chi}^{\prime}_i, \nonumber \\
{\cal L}_{F}&=& \frac{1}{3}[
3 \lambda_{1} \epsilon F_L \tilde{L} \tilde{L}+
\lambda_{2} \epsilon ( 2F_{L}\eta + F_{ \eta}\tilde{L}) \tilde{L}+
f_{1} \epsilon (F_{ \rho} \chi \eta+ \rho F_{ \chi} \eta+ \rho \chi F_{ \eta}) \nonumber \\
&+&
f^{\prime}_{1} \epsilon (F_{ \rho^{\prime}} \chi^{\prime} \eta^{\prime}+ 
\rho^{\prime} F_{ \chi^{\prime}} \eta^{\prime}+ 
\rho^{\prime} \chi^{\prime} F_{ \eta^{\prime}})
+ \kappa_{1}(F_{Q_1}\eta^{\prime}\tilde{u}^{c}_{i}+\tilde{Q}_1F_{\eta^{\prime}}
\tilde{u}^{c}_{i}+\tilde{Q}_1\eta^{\prime}F_{u_{i}}) \nonumber \\
&+& 
\kappa_{2}(F_{Q_1}\rho^{\prime}\tilde{d}^{c}_{i}+\tilde{Q}_1F_{\rho^{\prime}}
\tilde{d}^{c}_{i}+\tilde{Q}_1\rho^{\prime}F_{d_{i}}) 
+  \kappa_{3}(F_{Q_1}\chi^{\prime}\tilde{J}^{c}+
\tilde{Q}_1F_{\chi^{\prime}}\tilde{J}^{c}+\tilde{Q}_1\chi^{\prime}F_{J}) 
\nonumber \\
&+& \kappa_{4}(F_{Q_{\alpha}}\eta 
\tilde{d}^{c}_{i}+\tilde{Q}_{\alpha}F_{\eta}\tilde{d}^{c}_{i}+
\tilde{Q}_{\alpha}\eta F_{d_{i}}) 
+ \kappa_{5}(F_{Q_{\alpha}}\rho\tilde{u}^{c}_{i}+
\tilde{Q}_{\alpha}F_{\rho}\tilde{u}^{c}_{i}+\tilde{Q}_{\alpha}\rho F_{u_{i}}) 
\nonumber \\
&+& \kappa_{6}(F_{Q_{\alpha}}\chi 
\tilde{j}^{c}_{\beta}+
\tilde{Q}_{\alpha}F_{\chi}\tilde{j}^{c}_{\beta}+\tilde{Q}_{\alpha}\chi 
F_{j_{\beta}}) 
+ \kappa_{7}(F_{Q_{\alpha}}\tilde{L}
\tilde{d}^{c}_{i}+\tilde{Q}_{\alpha}F_{L}\tilde{d}^{c}_{i}+
\tilde{Q}_{\alpha}\tilde{L}F_{d_{i}}) \nonumber \\
&+& \xi_{1}(2F_{d_{i}}\tilde{d}^{c}_{j}
\tilde{u}^{c}_{k}+\tilde{d}^{c}_{i}\tilde{d}^{c}_{j}F_{u_{k}}) 
+ \xi_{2}(2F_{u_{i}}
\tilde{u}^{c}_{j}\tilde{j}^{c}_{\beta}+\tilde{u}^{c}_{i}\tilde{u}^{c}_{j}
F_{j_{\beta}}) \nonumber \\
&+&\xi_{3}(F_{d_{i}}\tilde{J}^{c}
\tilde{j}^{c}_{\beta}+
\tilde{d}^{c}_{i}F_{J}\tilde{j}^{c}_{\beta}+
\tilde{d}^{c}_{i}\tilde{J}^{c}F_{j_{\beta}})+\lambda_{4} \epsilon (F_{L}\chi 
\rho+ \tilde{L}F_{\chi}\rho+ \tilde{L} \chi F_{\rho})], \nonumber \\
{\cal L}_{ll \tilde{l}}&=&- \frac{ \lambda_{1}}{3} \epsilon 
(LL\tilde{L}+\tilde{L}LL+L\tilde{L}L); \,\
{\cal L}_{llH}=- \frac{1}{3}
\lambda_{2} \epsilon LL\eta , \nonumber \\
{\cal L}_{l \tilde{l} \tilde{H}}&=&- \frac{1}{3} 
\lambda_{2} \epsilon ( \tilde{L}L\tilde{\eta}+L\tilde{L}\tilde{\eta}); \,\
{\cal L}_{l\tilde{H}H}=- \frac{ \lambda_{4}}{3} \epsilon 
(L \tilde{\chi} \rho+L \chi \tilde{\rho}), \nonumber \\  
{\cal L}_{H \tilde{H} \tilde{H}}&=&- \frac{1}{3}[ 
f_{1} \epsilon ( \tilde{ \rho} \tilde{ \chi} \eta+ \rho \tilde{ \chi} 
\tilde{ \eta}+ \tilde{ \rho} \chi \tilde{ \eta})+
f^{\prime}_{1} \epsilon(\tilde{\rho}^{\prime}\tilde{\chi}^{\prime}\eta^{\prime}
+ \rho^{\prime} \tilde{ \chi}^{\prime} \tilde{ \eta}^{\prime}+ 
\tilde{ \rho}^{\prime} \chi^{\prime} \tilde{ \eta}^{\prime})
], \nonumber \\
{\cal L}_{qqH}&=&- \frac{1}{3}[ 
\kappa_{1} Q_1\eta^{\prime}u^{c}_{i}+
\kappa_{2} Q_1\rho^{\prime}d^{c}_{i}+
\kappa_{3} Q_1\chi^{\prime}J^{c}+
\kappa_{4} Q_{\alpha}\eta d^{c}_{i}+
\kappa_{5} Q_{\alpha}\rho u^{c}_{i}+
\kappa_{6} Q_{\alpha}\chi j^{c}_{\beta}], \nonumber \\
{\cal L}_{q \tilde{q} \tilde{H}}&=&- \frac{1}{3}[
\kappa_{1}(Q_1 \tilde{u}^{c}_{i}+ \tilde{Q}_1u^{c}_{i}) \tilde{ \eta}^{\prime}+
\kappa_{2}(Q_1 \tilde{d}^{c}_{i}+ \tilde{Q}_1d^{c}_{i}) \tilde{ \rho}^{\prime}+
\kappa_{3}(Q_1 \tilde{J}^{c}+ \tilde{Q}_1J^{c}) \tilde{ \chi}^{\prime} 
\nonumber \\
&+&
\kappa_{4}(Q_{\alpha} \tilde{d}^{c}_{i}+ \tilde{Q}_{\alpha}d^{c}_{i}) 
\tilde{ \eta}+
\kappa_{5}(Q_{\alpha} \tilde{u}^{c}_{i}+ \tilde{Q}_{\alpha}u^{c}_{i}) 
\tilde{ \rho}+
\kappa_{6}(Q_{\alpha} \tilde{j}^{c}_{\beta}+ \tilde{Q}_{\alpha}j^{c}_{\beta}) 
\tilde{ \chi}], \nonumber \\
{\cal L}_{lq \tilde{q}}&=&- \frac{ \kappa_{7}}{3}(Q_{\alpha}\tilde{d}^{c}_{i}+
\tilde{Q}_{\alpha}d^{c}_{i})L; \,\
{\cal L}_{ \tilde{l}qq}=- \frac{ \kappa_{7}}{3} Q_{ \alpha} \tilde{L} d^{c}_{i}, \,\
{\cal L}_{\tilde{l}HH}=- \frac{ \lambda_{4}}{3} \tilde{L} \chi \rho , 
\nonumber \\
{\cal L}_{qq \tilde{q}}&=&- \frac{1}{3}[
\xi_{1}(d^{c}_{i}d^{c}_{j}\tilde{u}^{c}_{k}+\tilde{d}^{c}_{i}d^{c}_{j}u^{c}_{k}
+d^{c}_{i}\tilde{d}^{c}_{j}u^{c}_{k} )+
\xi_{2}(u^{c}_{i}u^{c}_{j}\tilde{j}^{c}_{\beta}+
\tilde{u}^{c}_{i}u^{c}_{j}j^{c}_{\beta}+u^{c}_{i}\tilde{u}^{c}_{j}
j^{c}_{\beta}) \nonumber \\
&+&
\xi_{3}(d^{c}_{i}J^{c}\tilde{j}^{c}_{\beta}+
\tilde{d}^{c}_{i}J^{c}j^{c}_{\beta}+d^{c}_{i}\tilde{J}^{c}j^{c}_{\beta})]. 
\nonumber \\
\label{HMT}
\end{eqnarray}


\end{document}